\documentclass[aps,superscriptaddress,floats,showpacs,floatfix,twocolumn]{revtex4}

\usepackage{amsmath}
\usepackage{amsfonts}
\usepackage{amssymb}
\usepackage{amsthm}
\usepackage{bm}
\usepackage{color}
\usepackage{xcolor}

\usepackage{graphicx}

\usepackage{algorithm}
\usepackage[noend]{algpseudocode}

\usepackage{dsfont}

\usepackage[colorlinks=true,linkcolor=blue,citecolor=magenta]{hyperref}

\newcommand{\cs}[1]{\textcolor{blue}{#1}}

\newcommand{\R}{\mathbb R}

\newcommand{\ep}{\varepsilon}
\newcommand{\Qnw}{Q^{\mathrm{nw}}}
\newcommand{\Qdm}{Q^{\mathrm{dm}}}

\newcommand{\Pdm}{P^{\mathrm{dm}}}
\newcommand{\uh}{u^{\mathrm{H}}}

\begin{document}
\title{Lagrangian coherent sets in turbulent Rayleigh--B\'enard convection}

\author{Christiane Schneide}
\affiliation{Institut f\"ur Mathematik und ihre Didaktik, Leuphana Universit\"at L\"uneburg, D-21335 L\"uneburg, Germany}
\author{Martin Stahn}
\affiliation{Institut f\"ur Mathematik, Freie Universit\"at Berlin, D-14195 Berlin, Germany}
\author{Ambrish Pandey}
\affiliation{Institut f\"ur Thermo- und Fluiddynamik, Technische Universit\"at Ilmenau, D-98684 Ilmenau, Germany}
\author{Oliver Junge}
\affiliation{Zentrum Mathematik, Technische Universit\"at M\"unchen, D-85748 Garching, Germany}
\author{P\'eter Koltai}
\affiliation{Institut f\"ur Mathematik, Freie Universit\"at Berlin, D-14195 Berlin, Germany}
\author{Kathrin Padberg-Gehle}
\affiliation{Institut f\"ur Mathematik und ihre Didaktik, Leuphana Universit\"at L\"uneburg, D-21335 L\"uneburg, Germany}
\author{J\"org Schumacher}
\affiliation{Institut f\"ur Thermo- und Fluiddynamik, Technische Universit\"at Ilmenau, D-98684 Ilmenau, Germany} 
\affiliation{Tandon School of Engineering, New York University, New York, NY 11201, USA}
\date{\today}

\begin{abstract}
Coherent circulation rolls and their relevance for the turbulent heat transfer in a two-dimensional Rayleigh--B\'{e}nard convection model are analyzed. The flow is in a closed cell of aspect ratio four at a Rayleigh number ${\rm Ra}=10^6$ and at a Prandtl number ${\rm Pr}=10$. Three different Lagrangian analysis techniques based on graph Laplacians (distance spectral trajectory clustering, time-averaged diffusion maps and finite-element based dynamic Laplacian discretization) are used to monitor the turbulent fields along trajectories of massless Lagrangian particles in the evolving turbulent convection flow. The three methods are compared to each other and the obtained coherent sets are related to results from an analysis in the Eulerian frame of reference. We show that the results of these methods agree with each other and that Lagrangian and Eulerian coherent sets form basically a disjoint union of the flow domain. Additionally, a windowed time-averaging of variable interval length is performed to study the degree of coherence as a function of this additional coarse graining which removes small-scale fluctuations that cause trajectories to disperse quickly. Finally, the coherent set framework is extended to study heat transport.      
\end{abstract}	
\maketitle

\section{Introduction}

Thermal turbulent convection acts as one essential driving mechanism in many turbulent flows in nature spanning a wide range of examples starting from stellar interiors \cite{Miesch2005} via planetary cores \cite{Christensen2006} to atmospheric motion \cite{Stevens2005} and transport dynamics in lakes and ponds \cite{Bouffard2019}.  An idealized model of thermal convection is Rayleigh--B\'{e}nard convection (RBC), in which a fluid layer placed between two solid horizontal plates is uniformly heated from below and cooled from above~\cite{Chilla:EPJE2012}. This particular setting contains already many of the properties which can be observed in natural flows. One is the formation of large-scale coherent patterns when RBC is investigated in horizontally extended domains \cite{Hartlep2003, Hardenberg2008, Bailon2010, Emran2015, Stevens2018, Pandey:NatCommun2018}. These coherent sets, which have been detected in the Eulerian frame of reference, are termed turbulent superstructures as the characteristic horizontal scale extends the height of the convection layer. In thermal convection flows, they consist of convection rolls and cells that may, however be concealed in instantaneous velocity fields by turbulent fluctuations. Large-scale circulations in Rayleigh--B\'{e}nard convection exist also in smaller domains or cells and have been analyzed for example by proper orthogonal decomposition (POD) \cite{Bailon2011,Podvin2012,Verma2017} (see also the Appendix of Verma \cite{Verma2018} for POD in RBC). In the Eulerian frame, large-scale patterns show up prominently either after time averaging or as the primary POD modes, for both, temperature and velocity fields. This is illustrated in Figure  \ref{fig:field} where two coherent circulation rolls are present with narrow regions of upwelling hot and downwelling cold fluid.

At the core of the data-driven analysis in the Lagrangian frame of reference is the concept of a {\em coherent set} \cite{FrLlSa10,Froyland2013,Allshouse2015,Karrasch2017}, a region in the fluid volume that only weakly mixes with its surrounding and which often stays regularly shaped (non-filamented) under the evolution by the flow.  Such regions can be determined in two ways, by set-oriented \cite{DellnitzFroylandJunge2001,DellnitzJunge2002} or manifold-based methods (see \cite{Allshouse2015,Hadjighasem2017} for recent reviews of both concepts). The manifold-based approach comprises Lagrangian coherent structures (LCS), i.e., minimal curves in two dimensions and surfaces in the three-dimensional case that enclose coherent sets \cite{Haller2015}. This framework was extended recently to include weak diffusion across the manifold \cite{Haller2018}. 

Coherent sets were originally introduced based on transfer operators \cite{FrLlSa10,Froyland2013}. These are linear operators that evolve densities under the action of the flow. Coherent sets can be identified from the leading singular functions of this operator. More recently, in Ref.~\cite{Froyland2015} they have been characterized as sets which possess a minimal boundary-to-volume ratio for the entire flow duration. Different approaches have been introduced recently that make use of spatio-temporal clustering algorithms applied to Lagrangian trajectory data \cite{Froyland_Padberg_2015,Hadjighasem2016,Banisch2017,Schlueter2017,Padberg2017,Schneide2018}. These algorithms aim at identifying coherent sets as groups of trajectories that remain close to each other in the time interval under investigation. 

In this work, we will focus on the latter of these Lagrangian approaches. The three methods that we are going to apply characterize coherence via regularized linear operators that are directly approximated on the basis of the Lagrangian trajectory data in the convection flow. These are the (i) transfer operator which is regularized by a diffusion kernel \cite{Banisch2017}, the (ii) graph Laplacian operator that characterizes a network of Lagrangian trajectories in the flow \cite{Padberg2017} and the (iii) dynamic Laplacian which characterizes sets with minimal averaged boundary-to-volume ratio \cite{Froyland2015,FroylandJunge2018} or equivalently almost invariant sets of a time-dependent heat flow \cite{Karrasch2017}. In all cases, gaps in the discrete eigenvalue spectrum of the operator under consideration give an indication of the number of coherent sets. As will be seen, all three Lagrangian methods detect the same core regions of the large-scale circulation rolls as the coherent sets in which fluid particles remain together for the longest time.  With progressing time these coherent sets get increasingly smaller in their spatial extent as expected for turbulent flows. 

A second aspect of this work is therefore to extend these Lagrangian concepts and to perform the analysis on data which are averaged in time over a window of variable length. Similar to the Eulerian studies which were mentioned at the beginning of this introduction, we want to investigate coherence as a function of this additional coarse graining which removes small-scale fluctuations in the flow that typically cause a fast separation of Lagrangian trajectories that are initially close together.  

A final aspect of this work is to adapt the presented analysis directly to the transfer of heat. The presented Lagrangian methods can then be used to investigate heat coherence in RBC.

Here, we study RBC in a two-dimensional closed box of aspect ratio four. Note that the large- and small-scale quantities show similar scalings in two- and three-dimensional RBC~\cite{Schmalzl:EPL2004, Poel:JFM2013, Pandey:Pramana2016} for large Prandtl numbers. Therefore, very-long-time temporal evolution of the convective flow configurations has been studied in two-dimensional settings~\cite{Petschel:PRE2011, Pandey:PRE2018}. The objective of this work is to take such a simple two-dimensional turbulent flow at a moderate Rayleigh number and to demonstrate and compare the Lagrangian concepts and ideas.

In Sec.~\ref{sec:dataset} we introduce the numerical model and the data set. Section~\ref{sec:coherence} gives a further motivation for coherence. The Lagrangian methods which we will compare are introduced in Sec.~\ref{sec:methods} and applied to the convection data in Sec~\ref{sec:comparison}.  Heat coherence is discussed in Sec~\ref{sec:heat_coh} and we conclude in Sec.~\ref{sec:conclusion}.

\section{Rayleigh--B\'{e}nard convection flow}
\label{sec:dataset}
Conservation of mass, momentum, and internal energy lead to equations which govern the dynamics of RBC. In the Boussinesq approximation~\cite{Chilla:EPJE2012} they are given in a non-dimensional form by
\begin{align}
\frac{\partial {\bm u}}{\partial t} + {\bm u} \cdot {\bm \nabla} {\bm u} & = -\nabla p + \theta {\bm e}_{z} + \sqrt{\frac{\rm Pr}{\rm Ra}} {\bm \nabla}^2 {\bm u}, \label{eq:u} \\
\frac{\partial \theta}{\partial t} + {\bm u} \cdot {\bm \nabla} \theta & = u_z + \frac{1}{\sqrt{\rm Pr Ra}} {\bm \nabla}^2 \theta, \label{eq:T} \\
{\bm \nabla} \cdot {\bm u} & = 0 \label{eq:m},
\end{align}
where ${\bm u}=(u_x,u_z)$, $\theta$, and $p$ are the velocity, temperature deviation, and pressure fluctuation fields, respectively. Note that the temperature fluctuation from the linear conductive profile is related to the total temperature field $T$ as 
\begin{equation}
T(x,z,t) = T_{\rm bottom} - \frac{\Delta T}{H} z + \theta(x,z,t),
\label{Teq}
\end{equation}
where $T_{\rm bottom}$ is the temperature at the bottom plate. Equation \eqref{Teq} is given here in physical units. Equations~(\ref{eq:u}--\ref{eq:m}) were nondimensionalized using the height of the simulation domain $H$ of as the characteristic length scale, the free-fall velocity $u_f = \sqrt{\alpha g \Delta T H}$ as the characteristic velocity, and the temperature difference $\Delta T$ between the top and bottom plates as the characteristic temperature. The main governing parameters of RBC are the Rayleigh number ${\rm Ra}$ and the Prandtl number ${\rm Pr}$. The Rayleigh number signifies the strength of thermal driving force compared to dissipative forces, and the Prandtl number is the ratio of the kinematic viscosity and thermal diffusivity of the fluid. They are defined as
\begin{align}
{\rm Ra} & = \frac{\alpha g \Delta T H^3}{\nu \kappa}, \\
{\rm Pr} & =  \frac{\nu}{\kappa},
\end{align}
where $\alpha, \nu, \kappa$ are the thermal expansion coefficient, the kinematic viscosity, and the thermal diffusivity of the fluid, respectively. The acceleration due to gravity $g$ points downwards.	

We assume that the fields $u_x, u_z, \theta \in H^1(\Omega \times [0,\tau])$ with $\Omega = [0,L_x] \times [0,H]$ and total integration time $\tau$. Here, $H^1$ is the Sobolev space of square integrable functions with square integrable derivatives. Equations~(\ref{eq:u}--\ref{eq:m}) are solved using a pseudospectral solver {\sc Tarang}~\cite{Verma:Pramana2013} in a two-dimensional box of aspect ratio $\Gamma = L_x/H = 4$. Stress-free (or free-slip) boundary conditions for the velocity field are employed at all the walls. This implies that the corresponding normal velocity component and the normal derivative of the tangential velocity component vanish to zero, respectively. For the temperature field, isothermal (adiabatic) boundary conditions are applied in the vertical (horizontal) direction. To satisfy these boundary conditions, the temperature fluctuation and velocity components are expanded in sine and cosine basis functions. This results to
\begin{align}
u_x(x,z,t) & = \sum_{k_x,k_z} 4\hat{u}_x(k_x, k_z,t) \sin(k_x x) \cos(k_z z), \\
u_z(x,z,t) & = \sum_{k_x,k_z} 4\hat{u}_z(k_x, k_z,t) \cos(k_x x) \sin(k_z z), \\
\theta(x,z,t) & = \sum_{k_x,k_z} 4\hat{\theta}(k_x, k_z,t) \cos(k_x x) \sin(k_z z),
\end{align}
where ${\bm k} = (k_x,k_z)$ is the wave vector~\cite{Pandey:PRE2018} with $k_x = m \pi / L_x$ and $k_z = n \pi/H$; $m,n$ being integers. We perform direct numerical simulation for $\mathrm{Pr} = 10,$ $\mathrm{Ra} = 10^6$, and $\Gamma = 4$ using $513 \times 129$ uniformly spaced grid points. The time advancement is done using fourth-order Runge--Kutta method (RK4), and the fields are de-aliased using the 2/3 rule. We refer to \cite{Pandey:PRE2018, Verma:POF2015, Verma:Pramana2013} for more details on the numerical simulations.

The presented analyses require Lagrangian particle tracks which are evaluated together with the turbulent flow. Each individual tracer particle $i$ is advected in the velocity field corresponding to 
\begin{equation}
\frac{d {\bm x}_i}{dt}={\bm u} ({\bm x}_i,t)\,.  
\label{lag}
\end{equation}
We simulate $i=1\dots N$ particle trajectories with $N=5000$. Time integration is done again by the RK4 method. The interpolation of the velocity field on the particle position applies cubic splines. 

\begin{figure}
\centering
\includegraphics[width=0.47\textwidth]{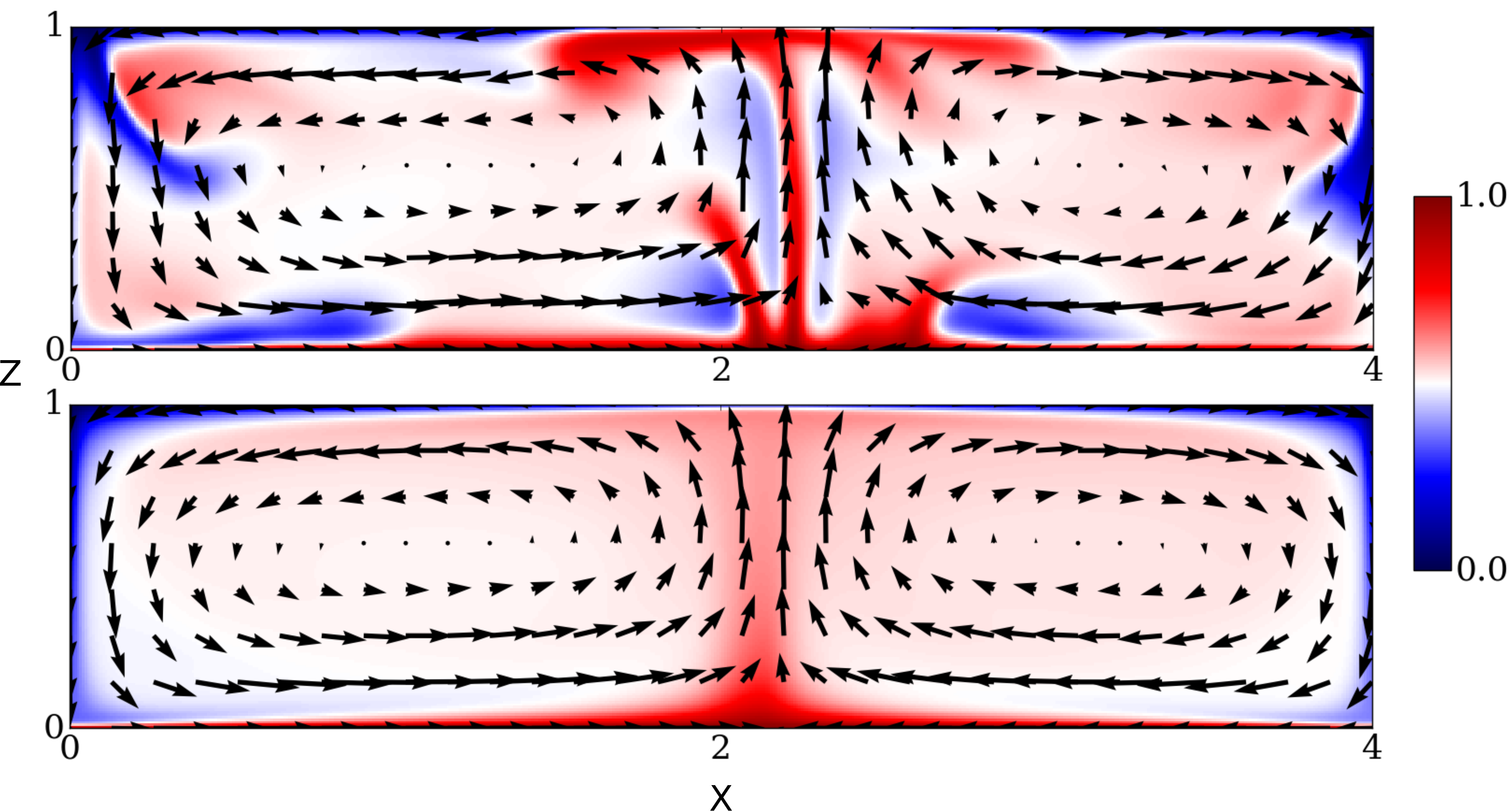}
\caption{(Color online) Two-dimensional RBC benchmark flow with Rayleigh number ${\rm Ra}=10^6$ and Prandtl number ${\rm Pr}=10$. Top: Instantaneous flow configuration exhibiting a pair of counter-rotating circulation rolls. Temperature contours are shown as colors and the velocity field is indicated by arrows. Bottom: Time-averaged flow configuration. Averaging is performed for the total duration of the time integration which is 500 free-fall time units.}
\label{fig:field}
\end{figure}

We start our simulation with random noise for velocity and temperature fields, ${\bm u}$ and $\theta$, as the initial condition and continue until a statistically steady state after 2000 free-fall times $t_f$ is reached. Here, $t_f = H/u_f$. The time-averaged flow structure exhibits a pair of counter-rotating circulation rolls as shown in the bottom panel of Figure~\ref{fig:field}. Hot fluid rises in the central region whereas cold fluid falls near the sidewalls.
The velocity and temperature fields at all the grid points are written out every~$0.1 \, t_f$.

\section{Lagrangian coherence in a turbulent flow}
\label{sec:coherence}

In the dynamical systems perspective, we consider the turbulent convection flow as a mapping in the state or phase space $\Omega \subset \R^2$. Let 
\begin{equation}
{\bm \Phi}_{t_{0}}^{t}\colon {\bm x}(t_0)\mapsto {\bm x}(t)={\bm \Phi}_{t_{0}}^{t}({\bm x}(t_0),t_0) 
\end{equation}
denote the flow map, which takes fluid particles from their initial location ${\bm x}_0={\bm x}(t_0)$ at time $t_{0}$ to their spatial location ${\bm x}(t)$ at time $t$ in correspondence with the velocity field ${\bm u}$ of the RBC flow. This mapping is given by the differential equations \eqref{lag} such that ${\bm \Phi}_{t_{0}}^{t}({\bm x}(t_0),t_0)$  solves the corresponding initial value problem. The advecting flow is simultaneously determined by solving the Boussinesq equations for the RBC flow. The phase space $\Omega$ is equipped with a reference measure $\mu$ and a sequence of non-singular time-dependent flow maps ${\bm \Phi}^{t_0+\Delta t}_{t_0}, \ldots , {\bm \Phi}^{t_0+n\Delta t}_{t_0+(n-1)\Delta t}$ for $n$ time steps $\Delta t$. We construct a single flow map via a successive application of flow maps over smaller time steps $\Delta t$, namely as ${\bm \Phi}={\bm \Phi}^{t_0+\Delta t}_{t_0}\circ \ldots \circ {\bm \Phi}^{t_0+n\Delta t}_{t_0+(n-1)\Delta t}$.

An overarching goal is to detect and locate slow mixing dynamical structures. These structures should be macroscopic in size and by ''slow mixing'' we have in mind a geometric mixing rate that is slower than $1/\Lambda$ where $\Lambda$ is the largest positive Lyapunov exponent that measures the exponential separation of initially infinitesimally distant neighboring particles. Thus, such slow mixing cannot be explained by local stretching, but is instead due to the way in which the dynamics acts globally. Following \cite{Froyland2015}, we wish to partition the state space $\Omega = \mathbb{A} \cup \mathbb{A}^C$ into a disjoint union of $\mathbb{A}$ and its complement $\mathbb{A}^C$ at initial time and $\Omega = \mathbb{B} \cup \mathbb{B}^C$ at final time such that we optimize the coherence ratio which can be thought of as 
\begin{equation}\label{eq:coh_quant}
\rho (\mathbb{A},\mathbb{B}) = \dfrac{ \mu (\mathbb{A} \cap {\bm \Phi}^{-1}(\mathbb{B}))}{\mu (\mathbb{A})} + \dfrac{\mu (\mathbb{A}^C\cap {\bm \Phi}^{-1}(\mathbb{B}^C))}{\mu(\mathbb{A}^C)}\,,
\end{equation}
and that quantifies how much of $\mathbb{A}$ ends up in $\mathbb{B}$ after applying the flow map ${\bm \Phi}$ (augmented by a small  random perturbation, to be precise) with respect to the reference volume of~$\mathbb{A}$ and similarly for $\mathbb{A}^C$ and $\mathbb{B}^C$. In a nutshell, Eq. \eqref{eq:coh_quant} quantifies in our example the area content of the subset that stays connected from a Lagrangian point of view.

We seek for an optimal 2-partition into slow mixing (or coherent) sets and the remainder which becomes rapidly well-mixed under the action of convective turbulence. Coherent means now that these sets are almost-invariant under the combined forward--backward dynamics of the flow map ${\bm \Phi}$ (including a small amount of diffusion for regularization). The exact version of \eqref{eq:coh_quant} and the connection to the transfer operator $\mathcal{P}$ and its singular values or functions can be found in \cite{Froyland2013}. 

\begin{figure}
\centering
\includegraphics[width=0.4\textwidth]{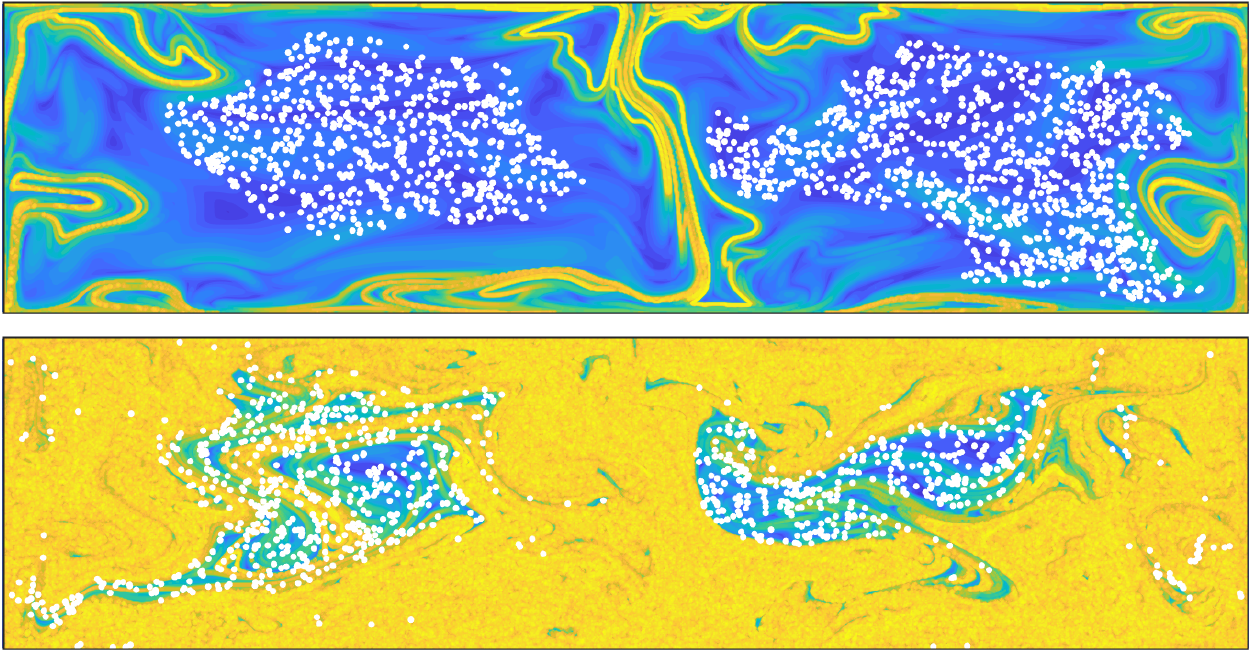}
\caption{(Color online) Finite-time Lyapunov exponent field computed in forward time over $20 t_f$ and $200 t_f$ in the top and bottom panel, respectively. Large values are indicated by yellow. Superimposed are particles (white dots) at initial positions ($t=2000 t_f$) which belong to two coherent sets as identified by the method described in section \ref{sec:nw} (see also Figures \ref{fig:ShortSlice} and \ref{fig:LongSlice}).
}
\label{fig:ftle}
\end{figure}

Central to manifold-based concepts of coherence are LCS \cite{Haller2015}. These are material surfaces that extremise a certain stretching or shearing quantity, such as measured by finite-time Lyapunov exponents (FTLEs) \cite{Haller2001}. Ridges in the FTLE field from a forward time computation highlight repelling LCS. In Figure \ref{fig:ftle} we show the forward time FTLE fields for the two different time windows considered in section \ref{ssec:experimental_comp}. While for the short time interval (top) isolated ridges can be observed, this is no longer the case for the long time interval (bottom). This makes it difficult to unambiguously define coherent sets from the FTLE ridges (see also \cite{Schneide2018} for related observations), and therefore, in the present work, we focus on set-based approaches. In Figure \ref{fig:ftle} we also show the results of a set-based computation (in this case using the minimum distance spectral trajectory clustering method introduced in section \ref{sec:nw}), which identifies coherent sets as the core regions of the large-scale circulation rolls that appear to be characterized by low FTLE values. We come back to this point in Sec. \ref{ssec:experimental_comp}.

\section{Graph Laplacian--based coherent structure detection}
\label{sec:methods}
All algorithms to be introduced work with the following set of Lagrangian data. Let the positions of $N$ particles ${\bm x}_i$ at $\mathcal{T}+1$ time instances $t_k$ be given, i.e., the trajectory data consists of
	\begin{equation} \label{eq:data}
		\left\{ {\bm x}_i(t_k) \in \R^d \; \middle| i \in \{1, \ldots , N\} \, , \, k \in \{ 0, \ldots , \mathcal{T}\} \right\} \subset \R^d,
	\end{equation}
where  ${\bm x}_i(t_{k+1})={\bm \Phi}_{t_{k}}^{t_{k+1}} {\bm x}_i(t_k)$, for  $i \in \{1, \ldots , N\}$ and  $k=0, \ldots , \mathcal{T}-1$. Trajectories in a coherent set will stay close to each other over a long time in contrast to trajectories not belonging to this specific coherent set. Diverging trajectories indicate filamentation of a set which has a large diffusive outward transport with respect to the dynamics. Based on this notion the algorithms evaluate the $\varepsilon$-neighborhood of the trajectories using distinct models of the diffusion process. If possible, this is followed by the construction of a rate matrix $Q$, which will be introduced in the following sections. The solution of an eigenvalue problem yields eigenvalues $\lambda_\ell$ which satisfy $0 = \lambda_1 \ge \lambda_2 \ge \ldots \ge \lambda_n$. The eigenvectors corresponding to the dominant eigenvalues (i.e.\ close to zero) are used to cluster the trajectory data into coherent sets.

\subsection{Distance spectral trajectory clustering}
\label{sec:nw}
The idea of the network--based analysis, published in \cite{Padberg2017}, is to interpret each Lagrangian trajectory $\{ {\bm x}_i(t_k)\}_k$ as node of a network consisting of $N$ nodes. A link between two nodes $\{{\bm x}_i(t_k)\}_k$ and $\{{\bm x}_j(t_k)\}_k$ is created if and only if the minimum distance of the two trajectories for at least one time instance $t_k$ is smaller than a pre-specified cutoff radius~$\varepsilon$. Thereby, network measures as, e.g., the node degree or local clustering coefficient can be used to distinguish coherent sets from incoherent flow \cite{Padberg2009, Donner2009, Banisch2019}. In order to partition the network into independent sets we use a version of the normalized cut approach \cite{ShiMalik2000}, which is based on spectral graph theory. This approach aims at maximizing the intra-cluster connectivity and simultaneously minimizing the inter-cluster connectivity. An approximate solution of the normalized cut problem can be achieved solving the generalized eigenvalue problem~\cite{ShiMalik2000}
	\begin{equation} \label{eq:GEP}
		Ly = \lambda Dy
	\end{equation}
where $L = D-A$ is the non-normalized graph Laplacian, $D$ is the degree matrix and $A$ is the adjacency matrix. Here, we vary the approach in the way that we use $L' = A-D$ and, for the sake of direct comparability with the second method, we state~\eqref{eq:GEP} in the equivalent form \footnote{Assuming $D$ has no zeros on the diagonal, i.e., every trajectory ``meets'' another at least once. If not, that is there is some~$i$ with $D_{ii}=0$, we can either delete the trajectory from the set completely, or if we want to keep it as an independent entity not connected to any other, we can set the $i$-th row of~$\Qnw$ equal to the $i$-th row of the identity matrix. The latter step produces a cluster containing only the $i$-th trajectory.},
	\begin{equation} \label{eq:EVPnw}
		\Qnw y = \lambda y, \quad\text{where } \Qnw = D^{-1}L'.
	\end{equation}
Note that the eigenvalues of~\eqref{eq:GEP} and~\eqref{eq:EVPnw} differ only in their signs. The multiplicity $m$ of the first eigenvalue $\lambda_1=0$ equals the number of connected components in the network. We construct the network such that the zero eigenvalue is simple. Then the number of eigenvalues $\lambda_2,\ldots\lambda_m$ close to $0$ determines the number of weakly linked subgraphs.
The eigenvectors corresponding to these eigenvalues are used to extract $m$ clusters. In this paper, this post-processing is done by the k-means algorithm \cite{Lloyd1982}. The corresponding pseudo-code is given in Algorithm~\ref{alg:min}.

\begin{algorithm}
\caption{Minimum distance spectral trajectory clustering}\label{alg:min}
\begin{algorithmic}[1]
\State Define adjacency matrix
\[
A_{ij} = \left\{
\begin{array}{ll}
1, & \|{\bm x}_i(t_k) - {\bm x}_j(t_k) \| \le \ep \\
    & \text{ for at least one } k\in\{0,\ldots,\mathcal{T}\} \text{ and }i\neq j, \\
0, & \text{otherwise}.
\end{array}
\right.
\]
\State Define degree matrix
\[
D_{ij} = \left\{
\begin{array}{ll}
\sum_{j'}A_{ij'}, & i=j \\
0, & \text{otherwise}.
\end{array}
\right.
\]
\State Define normalized Laplacian $\Qnw = D^{-1}(A-D)$.
\State Partition trajectories based on dominant eigenvectors \linebreak of $\Qnw$.
\end{algorithmic}
\end{algorithm}

We note that the adjacency matrix in Algorithm \ref{alg:min} could be constructed as a weighted matrix instead of the current binary one. One possibility is to measure the average distance between two trajectories
\begin{equation}
		\overline{d}({\bm x}_{i},{\bm x}_{j})=\frac{1}{\mathcal{T}+1}\sum_{k=0}^{\mathcal{T}}\lVert {\bm x}_{i}(t_{k})-{\bm x}_{j}(t_{k})\rVert,
\end{equation}
and set $A_{ij}=\overline{d}({\bm x}_{i},{\bm x}_{j})^{-1}$, $i\neq j$ as was suggested in~\cite{Hadjighasem2016}. In general, relaxing the binary structure of $A$ by introducing weights is a refinement of the dynamical information contained in $A$, and is used in the method discussed next.

\subsection{Time-averaged diffusion maps}
\label{sec:dm}

The theory of diffusion maps introduced by Coifman and Lafon in \cite{Coifman2006} has been successfully applied to a variety of nonlinear dimensionality reduction problems. In Ref.~\cite{Banisch2017} the framework of diffusion maps is used to analyze transport in dynamical systems and to find coherent sets solely based on possibly sparse or incomplete Lagrangian trajectory data. The idea is to introduce a diffusion process on the data points and detect points that can be reached more easily from one another by the diffusion process. This is done via the eigenvectors of this diffusion process (or operator) which provide intrinsic coordinates of the data set.

The time-averaged diffusion map algorithm from~\cite{Banisch2017} (called ``space-time diffusion map'' therein) proceeds as follows. Given the trajectory data as in~\eqref{eq:data}, the algorithm looks for tight bundles of trajectories. To achieve this, a diffusion map matrix $P_{t_k}$ is constructed at every time instance~$t_k$ by row-normalizing a similarity matrix~$K_{t_k}$. This matrix is constructed using a rotation invariant kernel which is given by 
\begin{equation}
k_{\delta}({\bm x},{\bm y}) = \exp\left(-\frac{\|{\bm x} -{\bm y}\|^2}{\delta}\right) \mathds{1}_{[0,r]}\,,
\end{equation}
with the characteristic function~$\mathds{1}_{[0,r]}$ introducing a cutoff at some radius~$r$. Hence, the similarity matrix~$(K_{t_k})_{ij} = k_\delta({\bm x}_i(t_k),{\bm x}_j(t_k))$ is only dependent on the Euclidean distances $\| {\bm x}_i(t_k) - {\bm x}_j(t_k) \|$ in $\R^2$ between the point pairs~${\bm x}_i(t_k), {\bm x}_j(t_k)$. The parameter $\delta$ can be seen as the strength (or duration) of the diffusion, and characterizes what will be considered close; hence its square root is a length scale of spatial resolution. In practice~$\delta$ together with the kernel function will determine a cutoff radius $r$, beyond which the similarity is negligibly small (here $r=\sqrt{2\delta}$). This is similar to $\ep$ in Sec.~\ref{sec:nw}, and allows for efficient numerical computation. Now, $P_{t_k}$ corresponds to a random walk (Markov chain) constructed on all existing data points (or particle positions) at $t_k$ with two points getting a higher jump probability the closer they are. Using time averaging,
	\begin{equation} \label{eq:Pdm}
		\Pdm = \frac{1}{\mathcal{T}+1} \sum_{k=0}^{\mathcal{T}} P_{t_k},
	\end{equation}
we obtain the space-time diffusion matrix $\Pdm$ which describes a Markov chain on all trajectories. We note, that averaging the matrices $P_{t_k}$ yields different results to using diffusion maps with the averaged distances $\overline{d}({\bm x}_{i},{\bm x}_{j})$ from above; and has a different interpretation (cf.~\cite{Banisch2017}).
	
Again, for the reason of comparability, we will work with the rate matrix
	\begin{equation} \label{eq:Qdm}
		\Qdm = \frac{1}{\delta}\big(\Pdm - I \big),
	\end{equation}
where $I$ denotes the identity matrix. By construction, the eigenvalues~$\lambda_\ell$ of~$\Qdm$ satisfy~$0 = \lambda_1 \ge \lambda_2 \ge \ldots \ge \lambda_n$. Just as in Sec.~\ref{sec:nw}, the dominant eigenvectors of~$\Qdm$ are now used to cluster the trajectory data into coherent sets.

\begin{algorithm}
\caption{Diffusion-map based analysis of Lagrangian data}\label{alg:dm}
\begin{algorithmic}[1]
\State Define kernel~$k_{\delta}({\bm x},{\bm y}) = \exp\big(-\|{\bm x}-{\bm y}\|^2 / \delta \big)\mathds{1}_{[0,r]} \big( \|{\bm x}-{\bm y}\| \big)$ with cutoff radius~$r$.
\State Define similarity matrices
\[
K_{t_k,ij} = k_{\delta} \big( {\bm x}_i(t_k), {\bm x}_j(t_k) \big).
\]
\State Define Markov matrices
\[
P_{t_k} = D_{t_k}^{-1}K_{t_k},
\]
with diagonal degree matrices $D_{t_k,ii} = \sum_{j'} K_{t_k,ij'}$.
\State Define time-averaged diffusion map matrix
\[
\Pdm = \frac{1}{\mathcal{T}+1} \sum_{k=0}^{\mathcal{T}} P_{t_k},
\]
and corresponding time-averaged diffusion-map Laplacian~$\Qdm = \frac{1}{\delta}\big(\Pdm - I \big)$.
\State Partition trajectories based on dominant eigenvectors of $\Pdm$ (which are the same as those of~$\Qdm$).
\end{algorithmic}
\end{algorithm}
We remark that the algorithm, as introduced here in Algorithm \ref{alg:dm}, corresponds to a particular choice of scaling in diffusion maps, the so-called~$\alpha=0$ case~\cite{Coifman2006, Banisch2017}. Originally in ref.~\cite{Banisch2017} a different scaling was used. We refer to that work for the details, and note only that for uniformly distributed data there is no practical difference. With the correct scaling, it is shown in Ref.~\cite[Thm.~3 and eq.~(21)]{Banisch2017} that if the number of trajectories goes to infinity,~$\Qdm$ converges to the so-called \emph{dynamic Laplacian} \cite{Froyland2015, FroylandKwok2017,Karrasch2017} whose eigenvectors characterize Lagrangian coherent sets. This makes the time-averaged diffusion-map method a consistent data-based approach for finding coherent sets.

\subsection{Discrete dynamic Laplacian}
\label{sec:ddL}
A third approach to the detection of Lagrangian coherent sets from sparse and possibly incomplete Lagrangian trajectory data has been developed in \cite{FroylandJunge2018}.  It is based on the geometric idea  \cite{Froyland2015} that Lagrangian coherent sets can be characterized by a small boundary-to-volume ratio: Whenever the length of the boundary of some advecting set $\bm\Phi_{t_0}^t(\mathbb{A})$, $\mathbb{A}\subset\Omega$, is  small in relation to its area consistently for all times $t$, diffusive transport of some passive scalar over its boundary (induced by small random perturbations to $\bm\Phi_{t_0}^t$)  will be small. Consequently, the coherence ratio \eqref{eq:coh_quant} of the pair $\mathbb{A},\bm\Phi_{t_0}^t({\mathbb A})$ will be large even in the presence of small perturbations to $\bm\Phi_{t_0}^t$.
  
Equivalently, cf. \cite{Karrasch2017}, we can characterize a Lagrangian coherent set as a material set which is almost-invariant under the flow of the Lagrangian diffusion equation \cite{Thiffeault2004} for some diffusive scalar quantity $S$. In Lagrangian coordinates, this equation is given by \cite{Karrasch2017}
\begin{equation}
\frac{\partial S}{\partial t}= \tilde\kappa_L {\bm \Delta}^t S ,
\label{eq:Lagrangian_heat_eq}
\end{equation}
with ${\bm \Delta}^t S = {\bm \nabla}\cdot\left({\cal D}({\bm x},t){\bm \nabla}S\right)$ and the dimensionless $\tilde\kappa_L\ll~1$. Here, the advection by the flow map that deforms a material set has been encoded in a Lagrangian eddy diffusivity (i.e.\ the inverse Cauchy-Green strain tensor) which is given by 
\begin{equation}
{\cal D}({\bm x},t)=D {\bm \Phi}_{t_{0}}^{t}({\bm x})^{-1}D{\bm \Phi}_{t_{0}}^{t}({\bm x})^{-\top}\,,
\end{equation}
where $D {\bm \Phi}_{t_0}^t$ is the Jacobian of the flow map.

In order to compute coherent sets via this approach, we first need to remove the time-dependence from the diffusion operator \mbox{${\bm \Delta}^t$}.  This can be achieved by time-averaging, i.e.\ by considering the operator
\begin{equation}
\bar{\bm \Delta} = \frac{1}{\mathcal{T}+1} \sum_{k=0}^{\mathcal{T}} {\bm \Delta}^{t_k},
\label{eq:dynamic_Lapacian}
\end{equation}
called the \emph{dynamic Laplacian} \cite{Froyland2015}.

In a second step, one considers the eigenproblem $\bar{\bm \Delta} v~=~\lambda v$ with appropriate boundary conditions (here, we have used Neumann boundary conditions).  Lagrangian coherent sets are then given by sublevel sets of the eigenvectors at the leading eigenvalues of $\bar{\bm \Delta}$, i.e.\ those closest to $0$ (cf.\ \cite{Dellnitz1999,DellnitzFroylandJunge2001,Froyland2015} for more details).

The dynamic Laplacian $\bar{\bm \Delta}$ in this eigenproblem can be discretized by standard finite element methods (FEM), leading to the generalized eigenproblem $\bar{K} v = \lambda M v$ with eigenvalues~$0 = \lambda_1 \ge \lambda_2 \ge \ldots \ge \lambda_n$. The stiffness matrix $\bar K$ is the average $\bar{K}=\frac{1}{\mathcal{T}+1} \sum_{k=0}^{\mathcal{T}} K^k$ of the stiffness matrices at each time $t_k$,
\begin{equation}
K_{ij}^k = -\int_\Omega \langle{\bm\nabla} \varphi^k_i, {\bm\nabla} \varphi^k_j\rangle\; dx.\label{eq:stiffness_matrix}
\end{equation}
Here, the functions $\varphi^k_i$ are the finite element basis functions on a triangulation of  the data points ${\bm x}_i(t_k)$. For the mass matrix $M$, it suffices to compute
\begin{equation}
M_{ij} = \int_\Omega \varphi^0_i \cdot \varphi^0_j\; dx
\label{eq:matrix_matrix}
\end{equation}
only at the initial time $t_0$. 

\begin{algorithm}
\caption{Dynamic Laplacian based analysis of Lagrangian data}\label{alg:dL}
\begin{algorithmic}[1]
\State Choose a finite element (e.g.,\ the piecewise linear triangular element).
\For{$k=0,\ldots,\mathcal{T}$}
\State  Construct a mesh of $\{{\bm x}_i(t_k), i=1,\ldots,N\}$.
\State  Compute $K^k$.
\EndFor
\State  Compute $\bar{K}=\frac{1}{\mathcal{T}+1} \sum_{k=0}^{\mathcal{T}} K^k$ and $M$.
\State Partition trajectories based on the dominant eigenvectors of $(\bar{K},M)$.
\end{algorithmic}
\end{algorithm}
A typical finite element in step 1 of Algorithm~\ref{alg:dL} is the linear triangular Lagrange element, i.e., the mesh in step 3 consists of triangles and the basis functions are the piecewise linear nodal basis functions. This computational mesh can be constructed as the Delaunay triangulation of the data points $\{{\bm x}_i(t_k), i=1,\ldots,N\}$ at each time step.  Whenever the data set has a very irregular hull, an $\alpha$-shape will be more appropriate. Typically, $\alpha$ is then chosen minimal such that the triangulation is connected (cf.\ \texttt{alphashape} in Matlab).  We refer to Ref. \cite{FroylandJunge2018} and to the documentation of the packages \href{https://github.com/gaioguy/FEMDL}{\texttt{FEMDL}} (Matlab) and \href{https://github.com/CoherentStructures/CoherentStructures.jl}{\texttt{CoherentStructures.jl}} (Julia), which are available from github, for more details and examples.

The triangulation induces a network of trajectories for each time step. There is an interpretation of this discretization of the dynamic Laplacian from a graph-Laplacian perspective, detailed in Appendix~\ref{app:dynLap_graphLap}.

Finally, we want to state that there is no direct link to some rate matrix $Q$ as in the other two approaches since the inverse of $M$ will be in general not sparse.

\section{Comparison of different Lagrangian methods}
\label{sec:comparison}
In order to compare the methods introduced above, we choose different perspectives. We compare the approaches first on a theoretical level (\ref{ssec:theoretical_comp}) and regarding their structure (\ref{ssec:structural_comp}). Finally we apply all these three methods to the same trajectory data set introduced in Sec. \ref{sec:dataset} and compare their results (\ref{ssec:experimental_comp}).

\subsection{Theoretical level}
\label{ssec:theoretical_comp}

We start by noting that~$\Qnw$ and~$\Qdm$ are so-called \emph{rate matrices}. A rate matrix~$Q$ has the properties $\sum_{j} Q_{ij}=0$ for all $i$ and~$Q_{ij} \ge 0$ for $i\neq j$. It defines a time-continuous Markov chain in the following sense. The process being in state~$i$ at current time, (i) will stay in~$i$ for a random amount of time~$\tau$, where $\tau$ is an exponentially distributed random variable with rate~$-Q_{ii}$, \footnote{A random variable is exponentially distributed with rate~$\rho$, if the probability that~$\tau > t$ for any~$t>0$ is~$e^{-\rho t}$. Equivalently, the probability density of~$\tau$ is given by~$\smash{f_{\tau}(t) = \rho^{-1}e^{-\rho t}}$.} and (ii) then will jump to the next state~$j$ randomly (and independently of~$\tau$) with probability~$Q_{ij} / |Q_{ii}|$. Thus, the larger the absolute value of~$Q_{ii}$, the faster the jump occurs on average. The quantity~$1/|Q_{ii}|$ is called the \emph{(average/expected) holding time} of state~$i$. 

We refer to Appendix~\ref{app:theoretical_comp} for further technical details, and only note here briefly that the jump processes $\Qnw$ and $\Qdm$ introduce jumps of mean length $\mathcal{O}(\ep)$ and~$\mathcal{O}(\sqrt{\delta})$, respectively, and this distance governs the finest scales they can resolve. Coherent sets below these scales can not be detected. The FEM-based discretization of the dynamic Laplacian has no explicit scale parameter, however the average diameter of the triangulation can implicitly be viewed as such.

\subsection{Structure}
\label{ssec:structural_comp}
The descriptions in Sec. \ref{sec:methods} suggest that the algorithms construct similar objects and have partially even similar steps. All three algorithms are by their very nature frame independent (as they only use mutual distances of trajectories), fast, and construct sparse matrices that encode the space-time behavior. They require the following three parameters as user input:

(1) A subset~$\Theta \subset \{0,\ldots, \mathcal{T}\}$ of all time instances at disposal, Not every coherent set might be present for the entire observation range of a fully non-autonomous flow, thus a natural choice is to restrict the time interval of consideration to~$\Theta = \{k_0,\ldots,k_1\}$, where~$0\le k_0 < k_1 \le \mathcal{T}$. Also, if the sampling time step $\max_k |t_k - t_{k-1}|$ is small compared to typical dynamic time scales, the data can be subsampled with respect to time without essentially altering the results.

(2) Proximity parameters $\delta$ and $\varepsilon$ (not for the FEM approach). These govern the minimal spatial scales on which the methods can detect coherence. In the FEM method this scale will implicitly be defined by the size of the elements resulting from the triangulation. For uniformly distributed Lagrangian particles the diameter of the resulting triangulation can be interpreted as the implicit closeness parameter.
		
(3) The number of clusters. Gaps in the spectrum of $\Qdm$, $\Qnw$ or $(\bar{K},M)$ can indicate natural choices for the number of clusters. Some arguments for this approach are collected in \cite{Denner2017}.

All methods discussed here are highly suitable for sparse and incomplete data (see \cite{Banisch2017}, in particular section~V therein, \cite{Padberg2017} and \cite{FroylandJunge2018}).

\subsection{Numerical}
\label{ssec:experimental_comp}
In the following, we discuss how the three methods compare to each other when applied to the turbulent convection flow data introduced in Sec. \ref{sec:dataset} for times $t>2000 t_f$. In view to the Lagrangian coherence, we will look at a short time interval of ${\cal T}=200$ steps and a long time interval of ${\cal T}=2000$ steps with $0.1 t_f$ per step. We thus start at $t=2000.0 t_f$ and end at $\tau=2020.0 t_f$ and $\tau=2200.0 t_f$, respectively. These time intervals are chosen since the average circulation (or eddy turnover) time of one roll was found to be $t\approx 20 t_f$, calculated using the maximum circumference of the roll covering half of the box and the root mean square velocity (see \cite{Pandey:NatCommun2018} for more details). Thus, the shorter (longer) time interval approximately corresponds to $1$ circulation ($10$ circulations). We initialize $N=5000$ uniformly distributed particles in order to analyze mass transport. The particles are advected using the snapshot files of the velocity field and for comparability all three algorithms are applied to the same trajectory data set. 

At this point we would like to seize the opportunity to explain some options for the choice of the similarity parameters $\delta$ and $\varepsilon$. Some general criteria for the choice of $\delta$ and $\varepsilon$ are (i) sparsity of $\Qdm$ and $\Qnw$ to achieve for example $5\%$ non zero entries,  (ii) a stable spectrum which implies that for a series of $\delta$, $\varepsilon$ the dominant spectrum of $\Qdm$ or $\Qnw$ does not vary qualitatively thus conserving relative distances and gaps of eigenvalues, (iii) edge density for the graph- or network-based approaches \cite{Donner2010}. 

Aside from those technical constraints, there are physical reasons that account for the turbulent convection flow under consideration. For Rayleigh--B\'enard convection that is considered here, we use the Nusselt number ${\rm Nu}$, a dimensionless number that quantifies the global turbulent heat transfer across the plane, and the expected nondimensional radius of a convection roll $r_{roll}$ to obtain bounds for both cutoff scales:
\begin{equation}\label{eq:epsbounds}
\delta_T=\dfrac{1}{2 {\rm Nu}} < \varepsilon < \dfrac{r_{roll}}{2}\,.
\end{equation}
We recall here that $\varepsilon = \sqrt{2\delta}$. The lower bound in \eqref{eq:epsbounds} is determined by the mean thickness of the thermal boundary layer, $\delta_T=0.028$ for the present Nusselt number of ${\rm Nu}=17.7$. Recall that all length scales are expressed in units of height $H$. The viscous boundary layer thickness can be determined for stress-free boundary conditions by a method that has been suggested by Petschel {\it et al.} \cite{Petschel2016}. They determined a so-called dissipation layer thickness from the intersection points (close to the bottom and top walls) of the line- and time-averaged profile of the kinetic energy dissipation rate with its plane mean. The analysis for the present flow gives a viscous boundary layer thickness of $\delta_v\approx 0.039$ which is slightly larger than the thermal boundary layer thickness. 

The dynamics further restricts the choice of both closeness parameters. In the short time interval, the trajectories cover only one circulation and very few Lagrangian particles will switch from the left roll to the right roll or vice versa (see Fig.~\ref{fig:field}). Thus, we have to take a small cut-off parameter in order to capture relevant dynamical structures. For the long time interval, the spectra of eigenvalues are typically not stable in case of a small closeness parameter. To satisfy these restrictions for both methods, we get bounds on $\varepsilon$ for the short and long time intervals of 
\begin{equation}\label{eq:epsbounds_intervals}
\varepsilon \in\left\{\begin{array}{ll} \left[0.03, 0.07\right]    &:\; \tau = 2020 t_f \, , \\
                                                        \left[0.085, 0.25\right]  &:\; \tau = 2200 t_f \, .
              \end{array}\right.
\end{equation}    
As we will see further below, our choices satisfy $\delta_T < \delta_v < \varepsilon$.

In the following we want to visualize the differences and similarities of the results of the three algorithms. In general, the coherent sets are expected to be larger for the network-based analysis compared to the time-averaged diffusion maps method. The reason for this behavior is that the network-based approach does not take into account how close trajectories pass by (as long as it is closer than the threshold $\varepsilon$) and how long they reside in the vicinity of each other, thus having a less pronounced and hence larger ``dynamic neighborhood''. As the strength of the diffusion decreases exponentially with increasing distance between particles  diffusion the transition from coherent structure to incoherent flow is more easily detected for diffusion maps. Consequently, the definition of coherence is more strict in the diffusion maps method compared to the network-based method and this will lead to smaller coherent sets.  

\subsubsection{Short time interval up to $\tau=2020 t_f$}
\begin{figure}
\begin{center}
\includegraphics[width=0.48\textwidth]{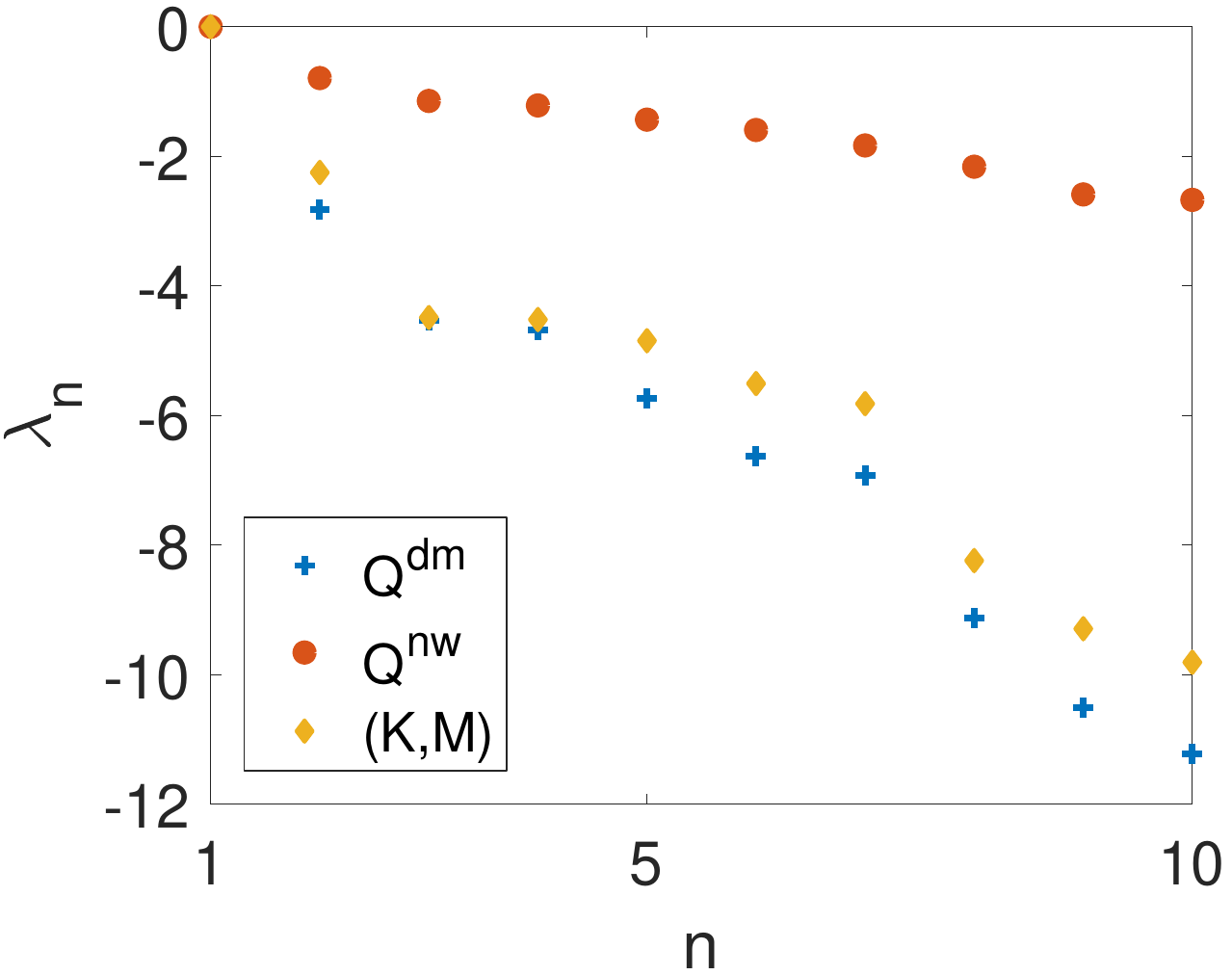}
\caption{(Color online) Eigenvalues of $2\varepsilon^{-1}\Qnw$ (circles), $\Qdm$ (crosses) and $(\bar K,M)$ (diamonds). The leading $10$ eigenvalues of the matrices of each approach for the short time interval setting are shown.}
\label{fig:ShortEigs}
\end{center}
\end{figure}

\begin{figure}
\begin{center}
\includegraphics[width=0.4\textwidth]{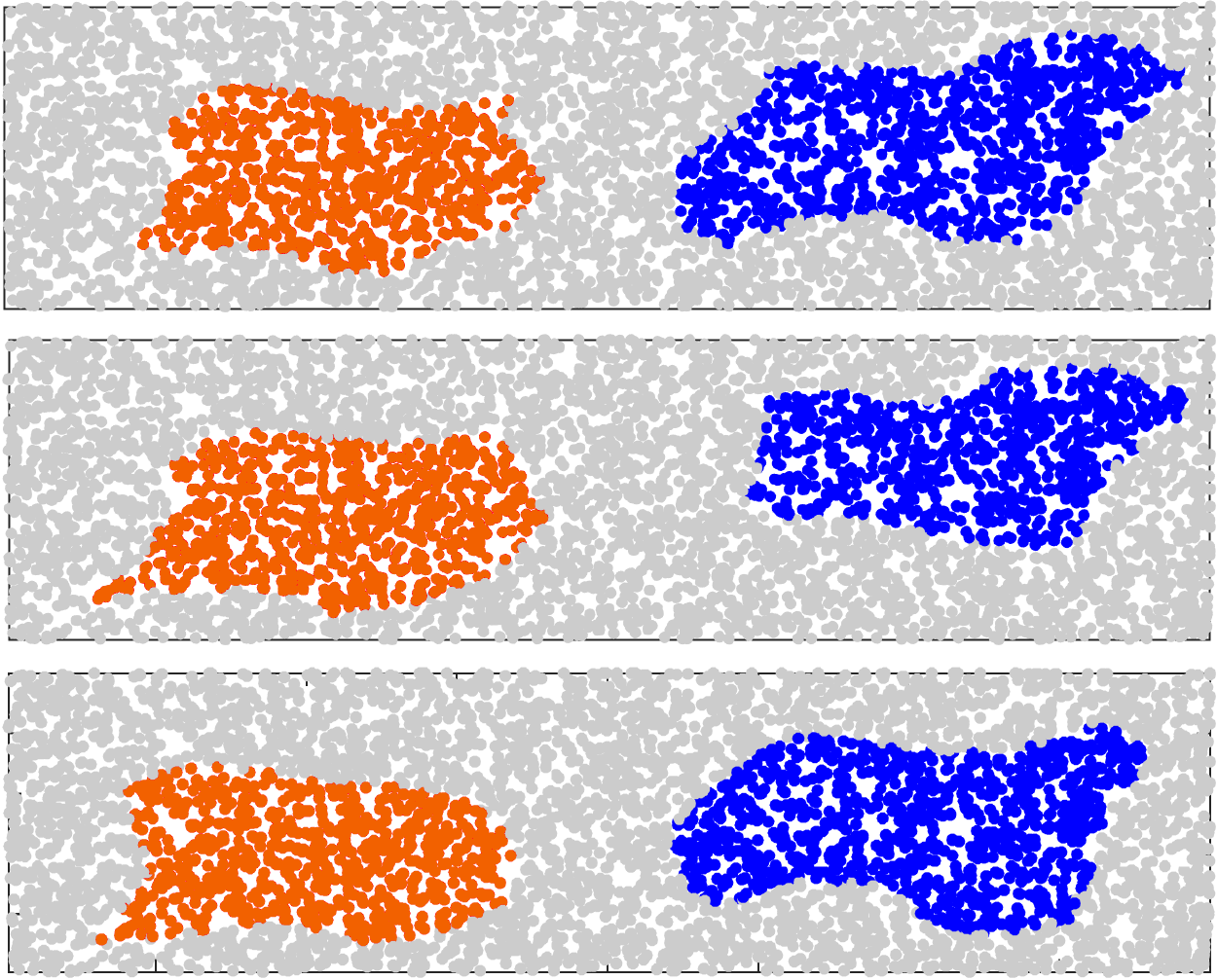}\\
\caption{(Color online) Clustering with respect to the second and fifth eigenvector of $\Qnw$(top) and of $\Qdm$ (center) and the second and third eigenvectors of $(\bar{K},M)$ (bottom). Particles in different coherent sets at time $t=2010t_f$ are colored according to $k$-means clustering.}
\label{fig:ShortSlice}
\end{center}
\end{figure}

\begin{figure}
\begin{center}
\includegraphics[width=0.4\textwidth]{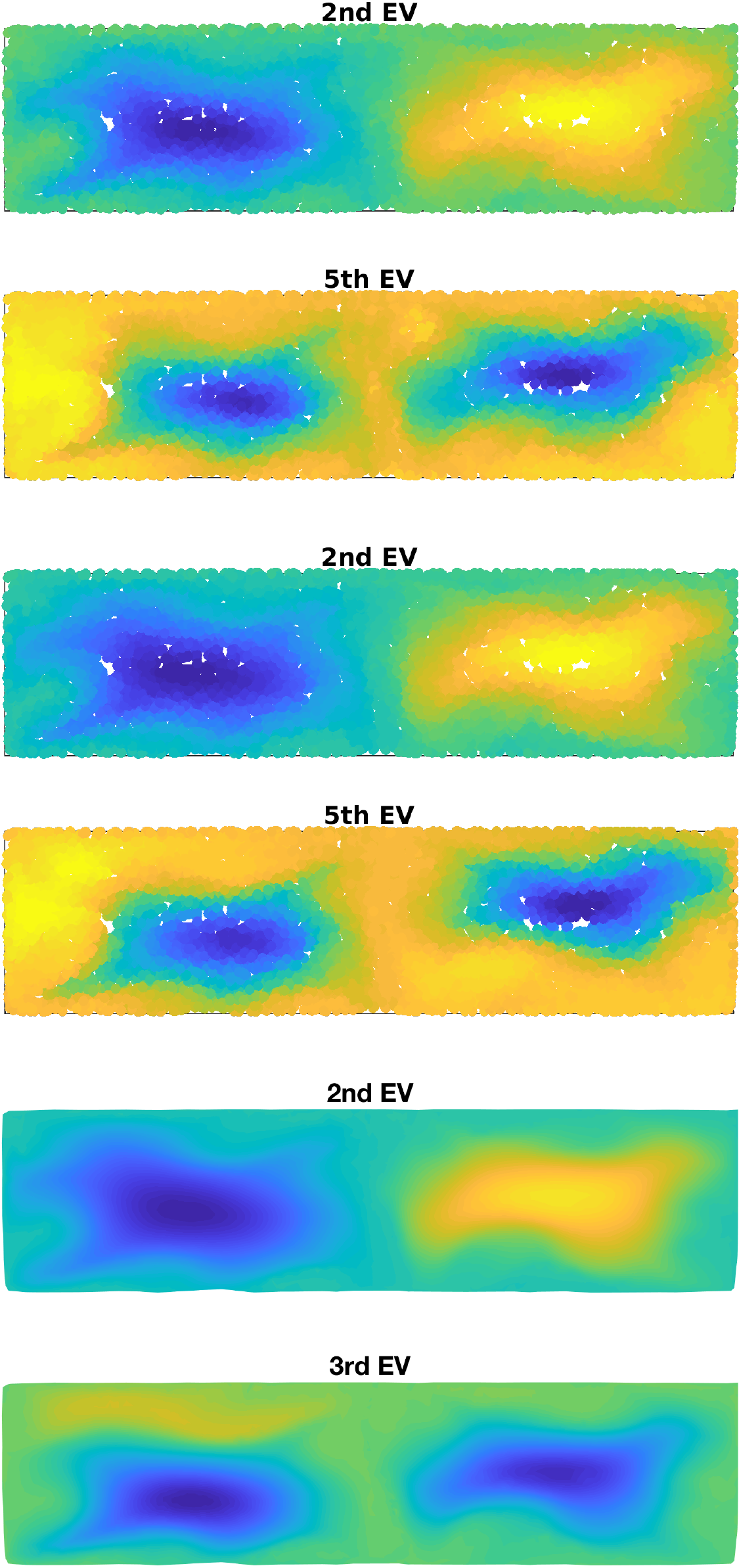}
\begin{picture}(1,1)
\put(-220,418){(a)}
\put(-220,344){(b)}
\put(-220,266){(c)}
\put(-220,196){(d)}
\put(-220,118){(e)}
\put(-220,42){(f)}
\end{picture}
\caption{(Color online) Particles colored according to eigenvectors of $\Qnw$ (a,b), $\Qdm$ (c,d) and $(\bar{K},M)$ (e,f) in the normalized range [-1,1]. Negative (positive) values are indicated by dark (bright) contours. The second and fifth (resp. third) eigenvectors at $t=2010t_f$.}
\label{fig:ShortEV}
\end{center}
\end{figure}
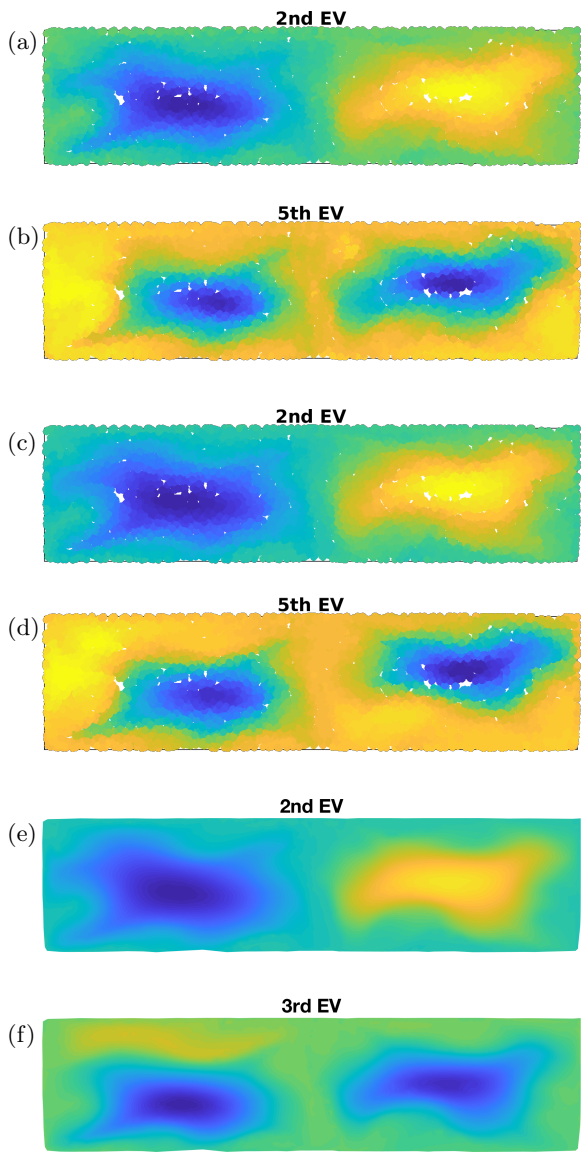

For the particular parameter choices $\delta = 0.002$ and $\varepsilon = 0.0632$, we now show the results for the short time span in Figures \ref{fig:ShortEigs}, \ref{fig:ShortSlice} and \ref{fig:ShortEV}. For visualization purposes we plot the eigenvalues of $2\varepsilon^{-1} \Qnw$.

The second eigenvector gives a separation of the left and right circulation rolls. The fifth eigenvector (resp. third for the FEM approach) gives a separation of the gyre cores and the background. We omit the plots for the third and fourth eigenvectors of $\Qnw$ and $\Qdm$ since they correspond to sub-partitions of the left and the right rolls only. They basically split each roll into halves, i.e., bottom and top or left an right for the third or fourth eigenvector, respectively. This occurrence can be explained in different ways. These (third and fourth) eigenvalues can be considered as higher multiplicities of the second eigenvalue interacting with the numerics. Furthermore these sub-partitions are still valid coherent sets which can be explained as follows. Within the selected time span the main motion is the rotation. However, even though most particles complete one rotation and the rotational speed varies with radius, this difference is not large enough to effectively separate particles from the inner (core) regions and the outer (background) regions. Therefore, the sub-partitions of the circulation rolls are more coherent than the gyre cores. As we will see later for the longer time span these sub-partitions do not occur, implying that they are then less coherent than the gyre cores.

\subsubsection{Long time interval up to $\tau=2200 t_f$}
\begin{figure}
\begin{center}
\includegraphics[width=0.48\textwidth]{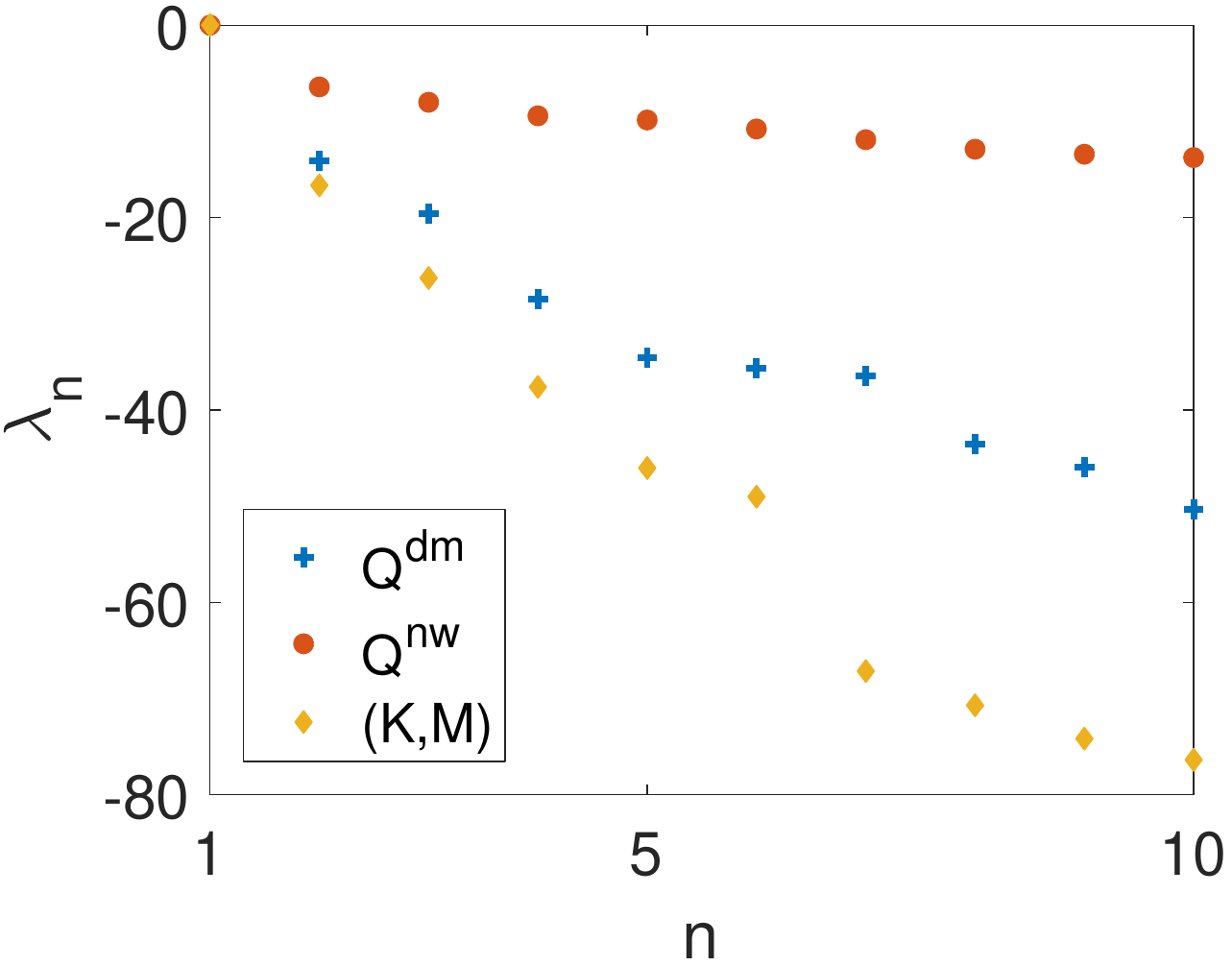}
\caption{(Color online) Eigenvalues of $2\varepsilon^{-1}\Qnw$ (circles), $\Qdm$ (crosses) and $(\bar K,M)$ (diamonds). Leading $10$ eigenvalues of the relevant matrices of each approach for the long time interval setting.}
\label{fig:LongEigs}
\end{center}
\end{figure}

In the following we visualize the results for the particular parameter choices of $\delta =0.005$ and $\varepsilon =0.1$ in Figures \ref{fig:LongEigs}, \ref{fig:LongSlice} and~\ref{fig:LongEV}. Again, for visualization purposes we plot the eigenvalues of $2\varepsilon^{-1} \Qnw$. We still get the separation of the left and the right side from the second eigenvector and the separation of the gyre cores and the background from the third eigenvector. We also note that the spectral gap is more prominent in the eigenvalue spectrum of $\Qdm$ compared to $\Qnw$. This could be advantageous in case of an unknown number of coherent sets. 
\begin{figure}
\begin{center}
\includegraphics[width=0.4\textwidth]{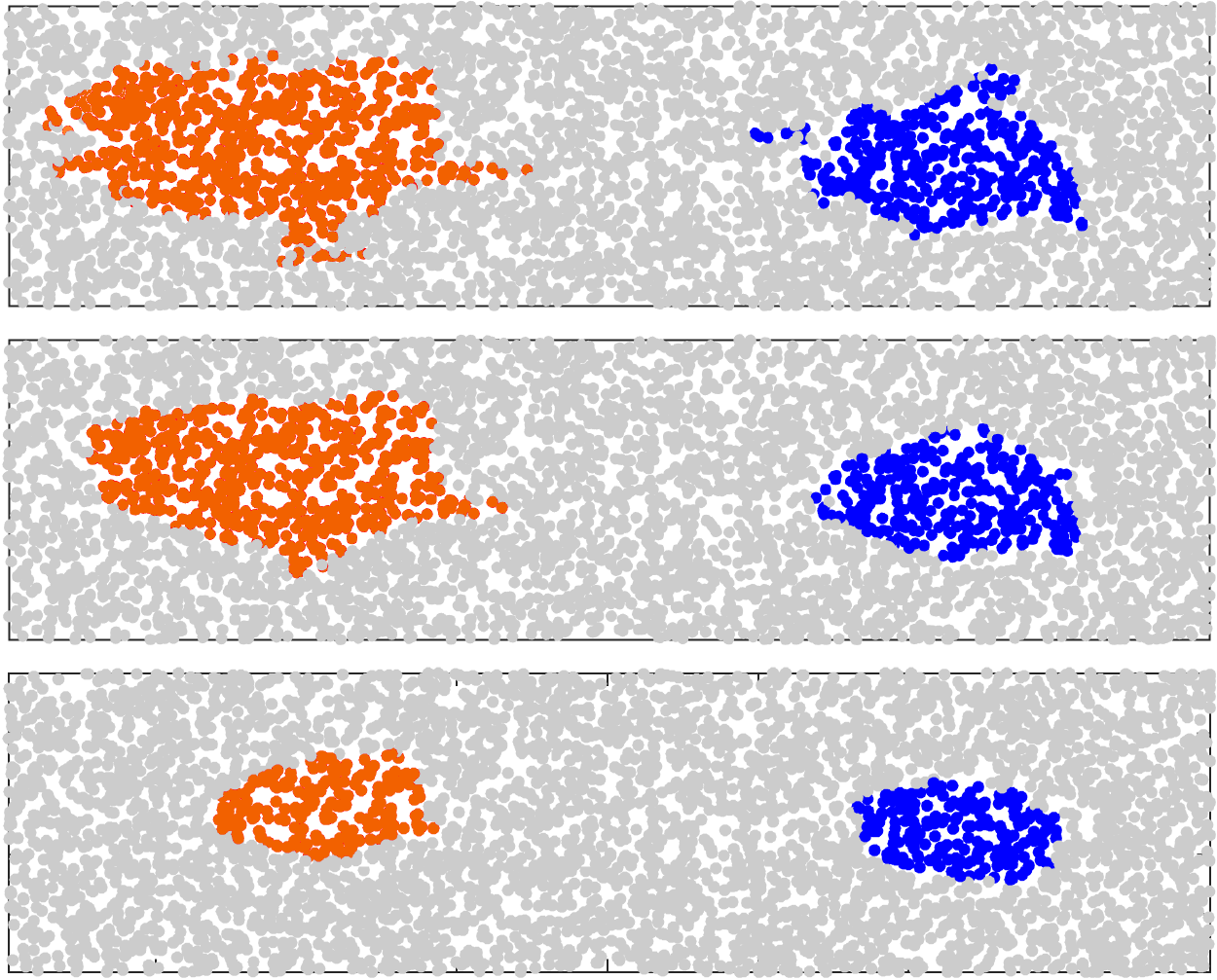}\\
\caption{(Color online) Clustering with respect to the second  eigenvector of $\Qnw$ (top), $\Qdm$ (center) and $(\bar{K},M)$ (bottom). Particles in different coherent sets at time $t=2100t_f$ are colored according to $k$-means clustering.}
\label{fig:LongSlice}
\end{center}
\end{figure}
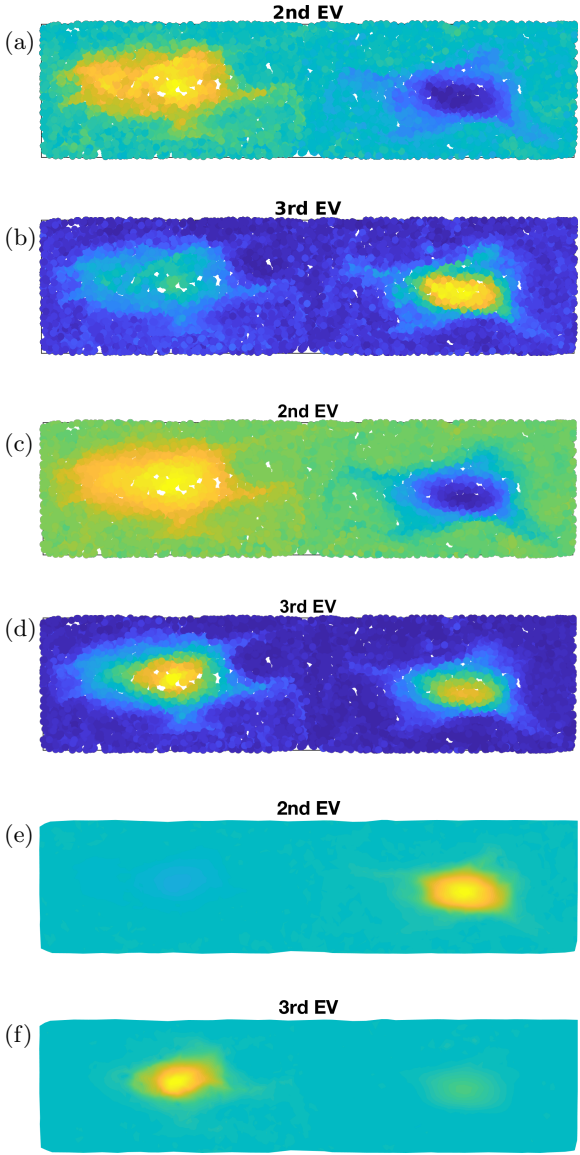
\begin{figure}
\begin{center}
\includegraphics[width=0.4\textwidth]{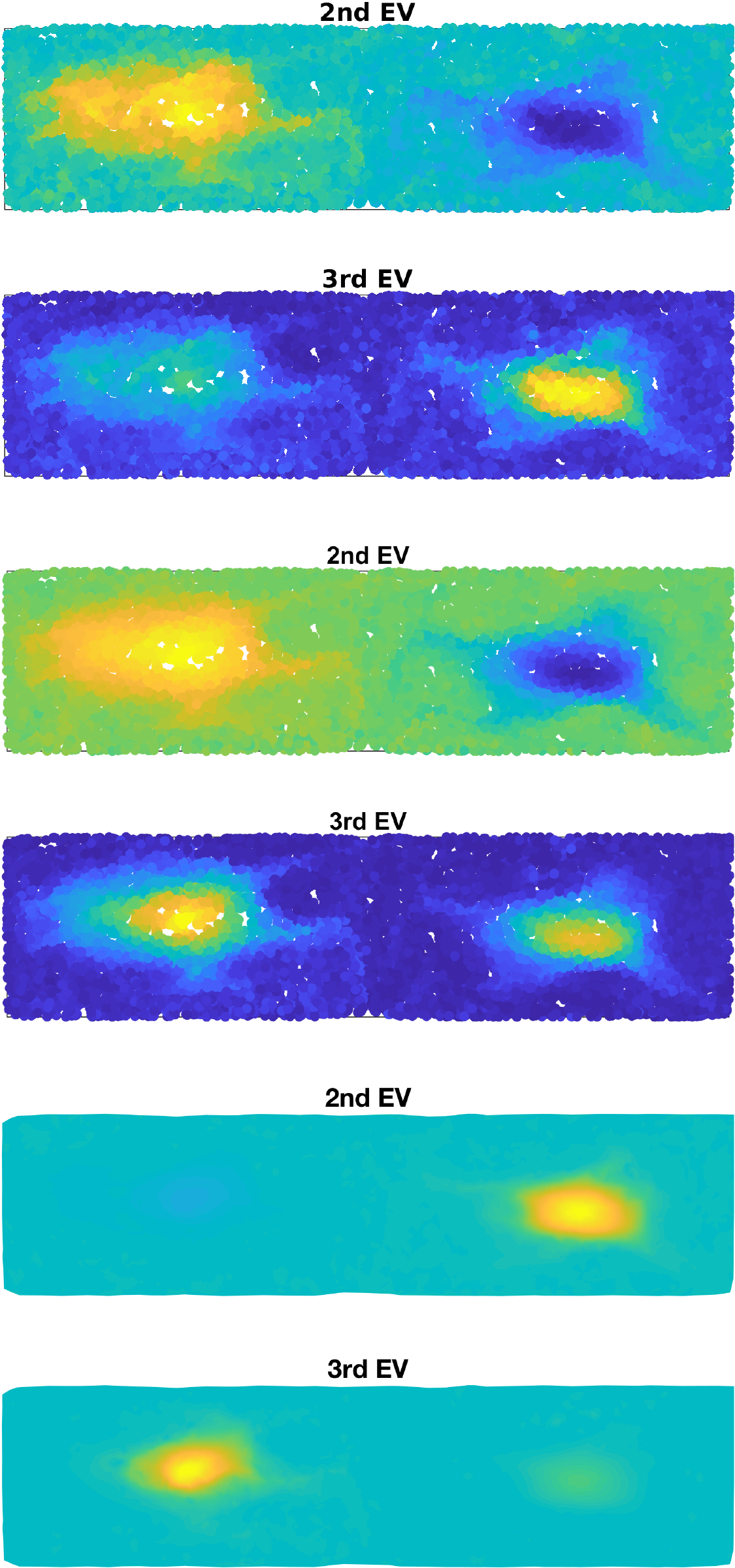}
\begin{picture}(1,1)
\put(-220,418){(a)}
\put(-220,344){(b)}
\put(-220,266){(c)}
\put(-220,196){(d)}
\put(-220,118){(e)}
\put(-220,42){(f)}
\end{picture}
\caption{(Color online) Particles colored according to eigenvectors of $\Qnw$ (a,b), $\Qdm$ (c,d) and $(\bar{K},M)$ (e,f) in the normalized range [-1,1]. Negative (positive) values are indicated by dark (bright) contours. The specific eigenvector is indicated at the top of each panel. The first non-trivial eigenvectors at $t=2100t_f$. }
\label{fig:LongEV}
\end{center}
\end{figure}

\subsubsection{Comparison of the three methods}
\begin{figure}
\begin{center}
\includegraphics[width=0.4\textwidth]{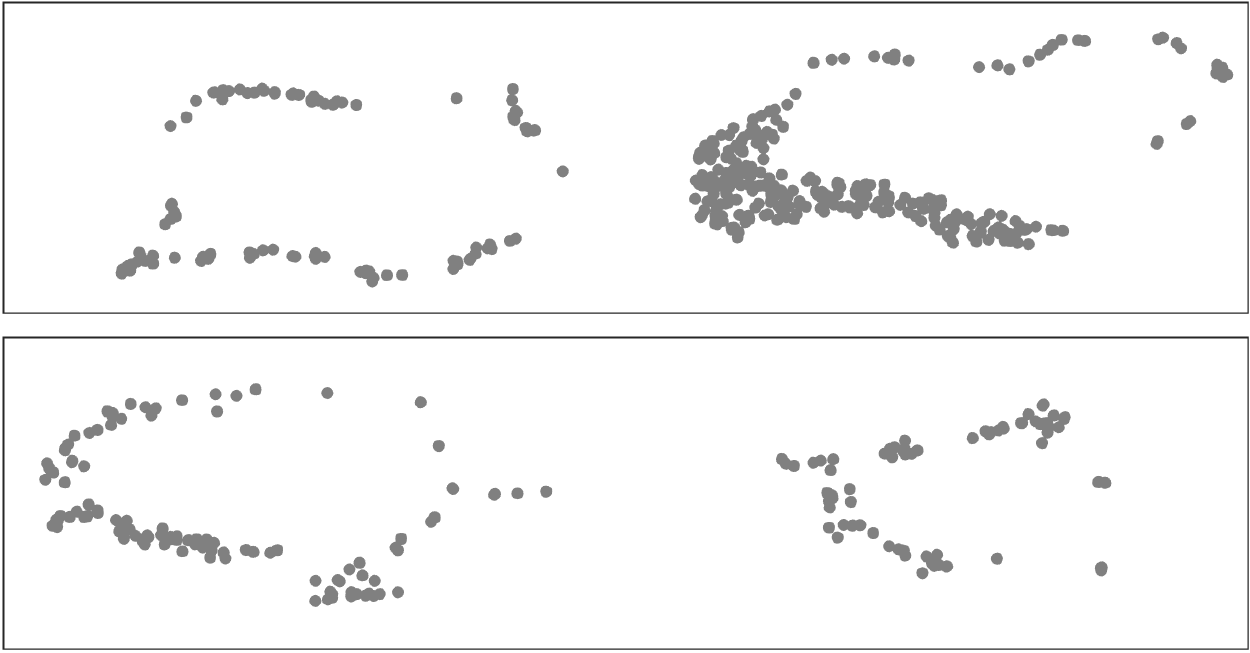}\\
\caption{Symmetrical difference of the clusters obtained from $\Qnw$ and $\Qdm$ at $t=2010 t_f$ (top) and at $t=2100 t_f$ (bottom).}
\label{fig:SliceComp}
\end{center}
\end{figure}

Figure~\ref{fig:SliceComp} compares the results by highlighting the trajectories that have been assigned differently by the network based and diffusion map method. This is done with a symmetrical difference of coherent sets $A^{\rm nw}$ and $A^{\rm dm}$, given by $(A^{\rm nw} \setminus A^{\rm dm}) \cup (A^{\rm dm} \setminus A^{\rm nw})$. For simplicity we only plot the symmetric differences for the $\Qnw$ and $\Qdm$ results. 
Furthermore, we observe a correlation between the distances between consecutive eigenvalues (see Fig.~\ref{fig:LongEigs}). 

In the short time setting the coherent sets resulting from the FEM method are comparable in size and shape to the ones detected by the other two methods.  For the long time interval, however, they are significantly smaller.  This might be explained by the original construction of the dynamic Laplacian via dynamic isoperimetry in Ref.~\cite{Froyland2015}: the method detects sets which keep a small boundary-to-volume ratio over the entire time interval of the flow evolution. It is thus a stricter criterion than the ones in the other two approaches.
\begin{figure}
\begin{center}
\includegraphics[width=0.4\textwidth]{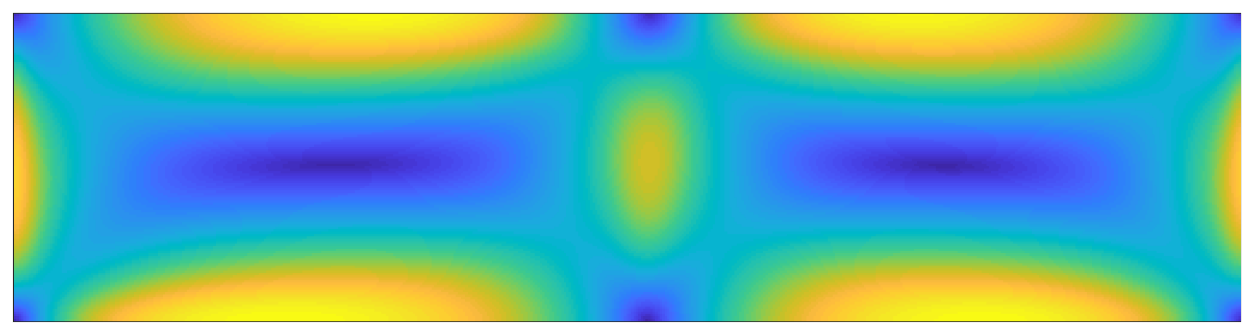}\\
\caption{(Color online) Magnitude of the time-averaged advecting velocity field in the plane as a filled contour plot. The averaging time is for 2000 $t_f$. The largest magnitudes are bright contours, the smallest ones are dark contours.}
\label{average}
\end{center}
\end{figure}

Figure \ref{average} displays the magnitude of the time-averaged velocity field. Small values can be identified in the elliptic centers of the counter-rotating rolls, the hyperbolic regions in the center of the top and bottom plates where Lagrangian particle pairs separate, and the four corners. The elliptic centers are the regions where the particles stay closely together for the longest time.

\subsubsection{Lagrangian particle advection in time-averaged flow}
\begin{figure}
\begin{center}
\includegraphics[width=0.48\textwidth]{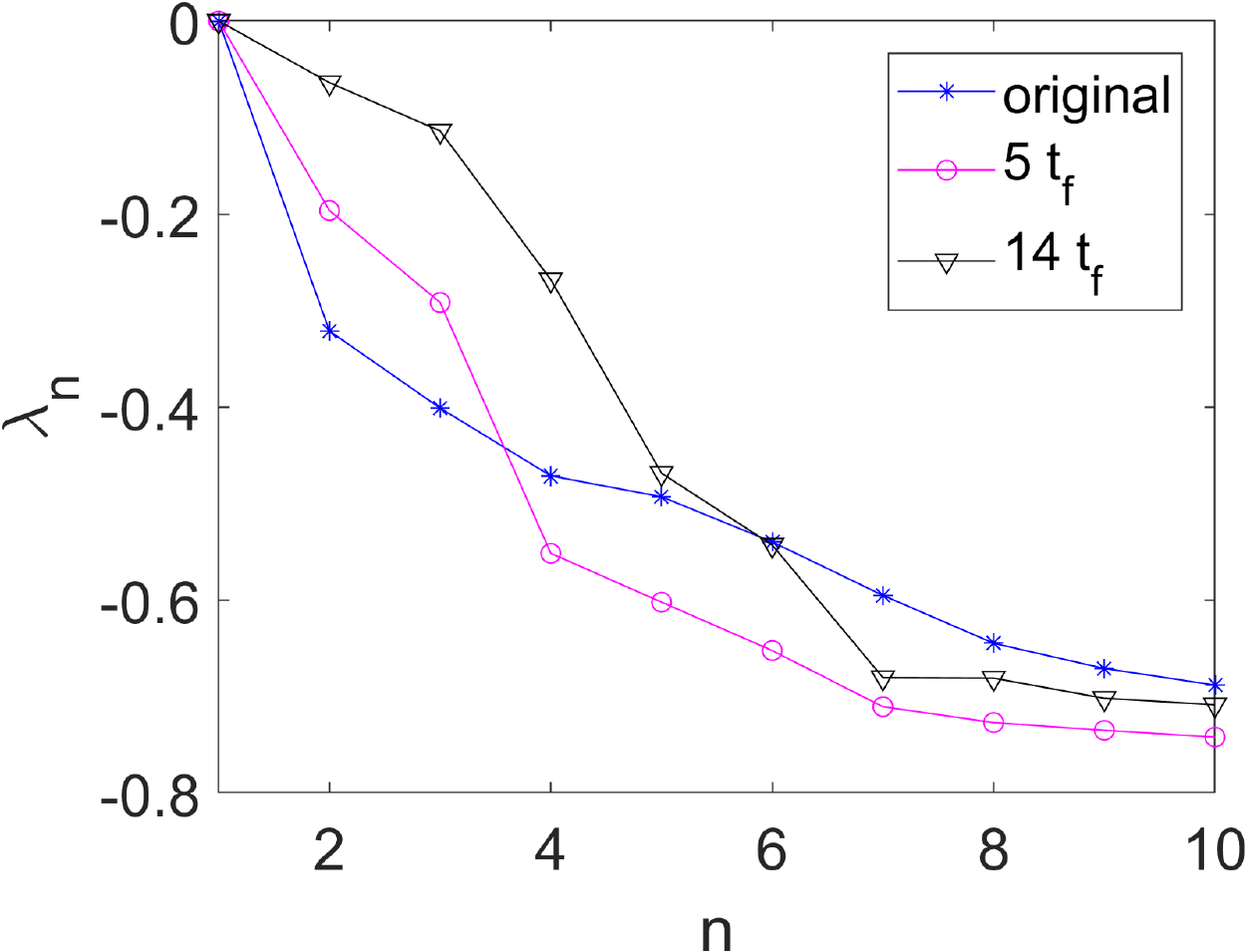}
\caption{(Color online) Effect of time-averaging in the Lagrangian analysis. Leading $10$ eigenvalues of $\Qnw$ of the original trajectories and time-averaged trajectories with window sizes $\Delta t = 5 t_f$ and $14 t_f$ for a time interval of $200.0 t_f$.}
\label{fig:TAEigs}
\end{center}
\end{figure}

\begin{figure}
\begin{center}
\includegraphics[width=0.4\textwidth]{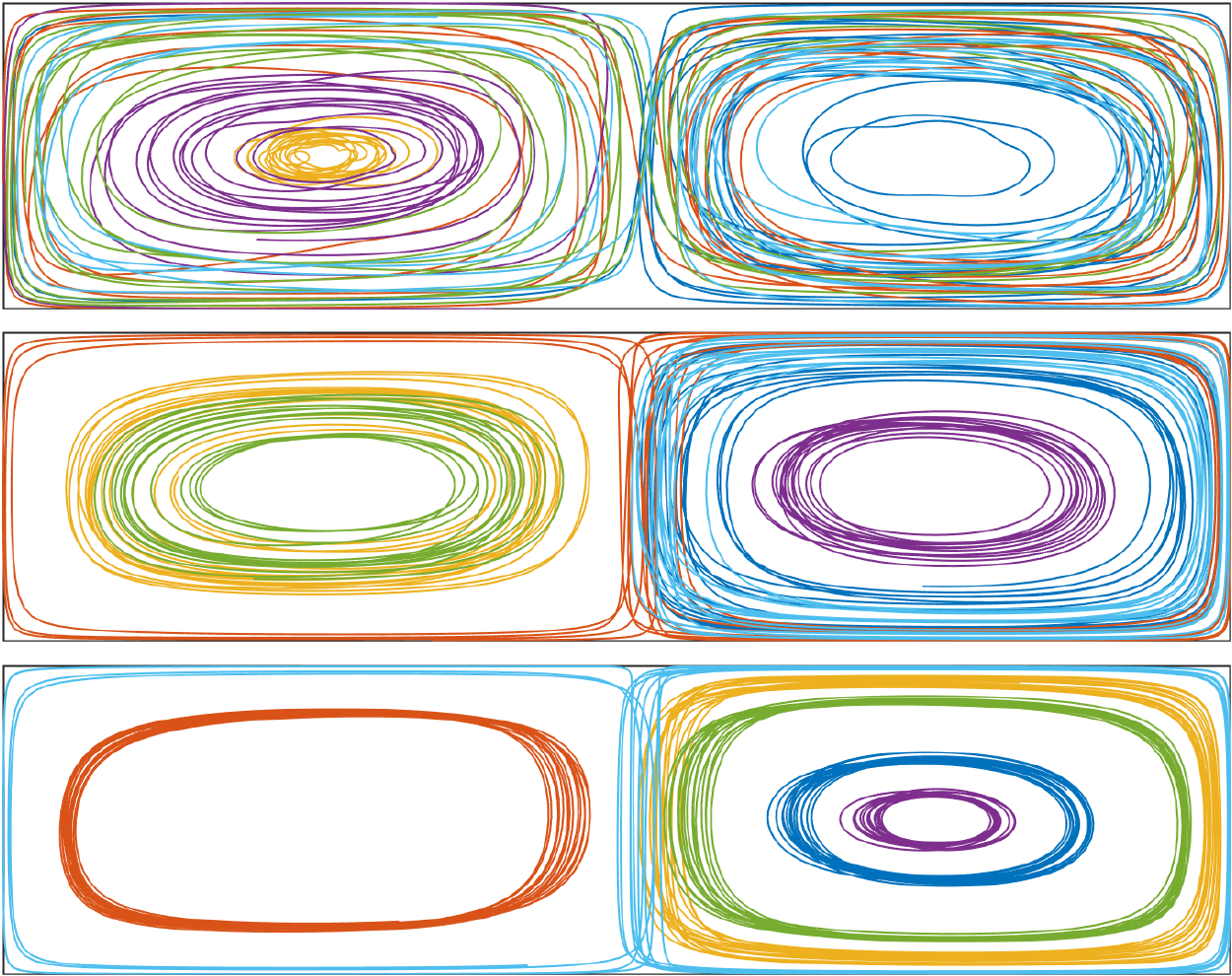}\\
\caption{(Color online) Comparison of trajectories for the original and time-averaged Lagrangian analysis. Top: trajectories of six random particles for a time interval of $200.0 t_f$. Middle (bottom): six time-averaged trajectories for $\Delta t = 5 t_f$ ($\Delta t = 14 t_f$).}
\label{fig:TATraj}
\end{center}
\end{figure}

In the Eulerian frame of reference, temporal averaging has to be performed in order to reveal the coherent large-scale patterns in the flow clearly~\cite{Pandey:NatCommun2018}. In the following, we adapt these ideas to the present Lagrangian analysis. We therefore carry out a time-averaging in the following sense
\begin{equation}
		\hat{{\bm x}}_{i}(t_{\hat{k}})=\frac{1}{\mathcal{T}_W}\sum_{j=\hat{k}}^{\hat{k}+\mathcal{T}_W-1} {\bm x}_{i}(t_{j}),
\end{equation}
where the number of time steps $\mathcal{T}_W$ depends on the chosen window size $\Delta t$ in free-fall time units. The time-averaged trajectories $\{ \hat{{\bm x}}_i(t_{\hat{k}})\}_{\hat{k}}$ are assembled such that they represent the same time interval of length $200 t_f$, i.e., $\hat{k}$ depends on~$\Delta t$. Here, we apply the minimum distance spectral trajectory clustering method (\ref{sec:nw}) to the time-averaged trajectories. Equivalent results can be achieved by applying the time-averaged diffusion maps method. The degree of coherence of the independent sets, interpreted as the size of the spectral gap, is improved by the time-averaging. Figure \ref{fig:TAEigs} shows the eigenvalues for the original setting, i.e., with no time-averaging, and for time-averaged trajectories with window sizes $\Delta t = 5 t_f$ and $14 t_f$ (which corresponds with ${\cal T}_W=50$ and 140 respectively). For $\Delta t = 5 t_f$ a prominent gap is visible between the third and fourth eigenvalues. This corresponds to three almost decoupled sets, i.e., dynamically independent flow regions, which are the two cores of the convection rolls and the background. With increasing window size transitions from one side of the domain to the other side, which occur rarely for individual trajectories in the original setting, are removed and the Lagrangian time-averaged trajectories $\{ \hat{{\bm x}}_{i}(t_{\hat{k}})\}_{\hat{k}}$ mostly remain in the initial flow region. Thereby, a segmentation into inner and outer cores is formed on both sides of the domain, corresponding to four almost decoupled sets. The spectral gap is shifted to appear between the fourth and fifth eigenvalues. This segmentation of the time-averaged trajectories is visualized in Figure~\ref{fig:TATraj}. The transition from three to four almost decoupled sets occurs around a window size of $\Delta t = 10 t_f$. We can thus conclude that an additional averaging enhances the long-term coherence of the sets significantly even for the present two-dimensional case for which the small-scale dispersion by turbulence remains moderate.

\section{Heat coherence in convection flow} 
\label{sec:heat_coh}

\subsection{Temperature from a passive scalar perspective}

The dynamical structure of heat transport can be analyzed from different perspectives. Although temperature is not a passive scalar, but a prognostic variable, once the evolution of the full system is known (i.e., the velocity $\bm u$ and total temperature $T$ fields are computed), we can view heat as a passive scalar with evolution governed by the advection-diffusion equation (see also eq.~\eqref{eq:T}) 
\begin{equation} 
\label{eq:heat-evol}
\frac{\partial T}{\partial t} = \frac{1}{\sqrt{\rm PrRa}} {\bm \nabla}^2 T - {\bm \nabla} \cdot({\bm u} T)
                                        =: - {\bm \nabla} \cdot {\bm \Psi},
\end{equation}
with boundary conditions that we have defined in Sec.~\ref{sec:dataset}. Here,~${\bm \Psi}$ denotes the total dimensionless heat flux vector and the average $\langle \Psi_z\rangle_{x,t}={\rm Nu}/\sqrt{\rm Ra Pr} $. The profile $T_0=1-z$ is the dimensionless linear equilibrium temperature profile (see also eq. \eqref{Teq}). Furthermore, $\Gamma_D$ and~$\Gamma_N$ are parts of the domain boundary, where prescribed constant temperature (i.e., Dirichlet conditon on $\Gamma_D$) and insulating wall (i.e., Neumann condition on $\Gamma_N$) boundary conditions are applied. To recall, an insulated side wall implies that the normal derivative vanishes, $\smash{\frac{\partial T}{\partial n}}=0$.	

This suggests that coherence with respect to heat could be analyzed by the approach from Sec.~\ref{sec:coherence}. The situation is more delicate, however, as the identification of ``heat packages'' with particles has to be done properly. One issue is, for instance, that heat is not a conserved quantity and can enter and exit at the top and bottom boundaries, $\Gamma_D$. Thus the boundary conditions in~\eqref{eq:heat-evol} have to be incorporated in the analysis. We start by briefly discussing a popular line of approaches that is however inappropriate for heat-coherence investigations.

In the so-called ``heatline'' approaches~\cite{Kimura1983,Costa2006,Speetjens2012,Pratt2016,Balasuriya2018}, the temperature field is considered as a concentration field driven by some velocity field~${\bm \uh}$. In order to get the correct flux,~${\bm \Psi} = {\bm \uh} T$ has to hold, giving
\begin{equation}
{\bm \uh}({\bm x},t) = {\bm u}({\bm x},t) - \dfrac{1}{\sqrt{\rm PrRa }} \dfrac{{\bm \nabla} T({\bm x},t)}{T({\bm x},t)}\,. \label{eq:heat-vel}
\end{equation}
Note that the system~\eqref{eq:heat-evol} is translation-invariant in the following sense: For any constant~$T_c\in\R$ and given solution~$T$ of~\eqref{eq:heat-evol}, the translated field~$T' = T + T_c$ solves~\eqref{eq:heat-evol} if the initial and Dirichlet boundary conditions are translated accordingly, i.e., $T'_{\Gamma} = T_{\Gamma} + T_c$ and~$T'_0 = T_0 + T_c$. This means intuitively, that the evolution does not change even if we express temperature with respect to an arbitrary reference value.

In contrast, the ``heat velocity'' field~${\bm \uh}$ changes under translation of the temperature field, giving completely different ``heat trajectories''. Nevertheless, by construction, the \emph{global temperature field}~$T$ is advected correctly by~${\bm \uh}$. This means, that the usefulness of the ``heat velocity field'' is restricted to considerations involving the global temperature distribution, but \emph{internal fluctuations} of heat and the \emph{internal transport} of heat are biased by the choice of reference temperature.
	
It should be remarked at this point that in \cite{Speetjens2012a, Speetjens2012} this problem is alleviated by splitting \emph{convective} and \emph{conductive} contributions of heat transport, where the translational invariance is taken up completely by the conductive part. To the remaining (convective) flux a velocity field~${\bm u}^{\mathrm{conv}} = {\bm \Phi}^{\mathrm{conv}} / T^{\mathrm{conv}}$ can be assigned, which then describes the convective heat transfer in a Lagrangian manner. However, as we would like to describe structures governing the entire heat transport (and not just its convective part), we take a different route in the following.

\subsection{Randomly evolving heat packages and induced transport}

The microscopic evolution of heat can be described by advection and stochastic fluctuations. Let us consider the stochastic differential equation  
\begin{equation} \label{eq:SDE_Langevin}
\dot{{\bm x}}(t) = {\bm u}({\bm x}(t),t) + \sqrt[4]{\frac{4}{\rm Pr Ra}}
{\bm \eta}(t),
\end{equation}
where ${\bm x}(t)$ denotes the random position of a particle driven by the drift~${\bm u}$ and by white noise~${\bm \eta}$ (i.e., $\mathrm{Cov}[{\bm \eta}(t),{\bm \eta}(s)] = \delta(t-s)$, where~$\delta$ denotes the Dirac delta distribution). The probability distribution~$p({\bm x},t)$ of~${\cs{\bm x}}(t)$ satisfies the Fokker--Planck equation
\begin{align}
\frac{\partial p}{\partial t} = \dfrac{1}{\sqrt{\rm PrRa}} {\bm \nabla}^2 p
                                  - {\bm \nabla}\cdot \left({\bm u} p \right).
\end{align}
If we replace~$p$ by $T$, this is identical with the evolution equation in~\eqref{eq:heat-evol}. Thus, scaling $T$ to have integral one (setting the total heat to be unity), we can express the evolution of heat as the evolution of the probability distribution of an ensemble of particles, evolving mutually independently with their motions governed by~\eqref{eq:SDE_Langevin}. This ensemble has initial distribution~$T_0$. Single particles can thus be viewed as ``heat packages'', each carrying one unit of heat.

We will illustrate some adaptations necessary according to the boundary conditions in the Rayleigh--B\'enard convection problem. Neumann zero boundary conditions naturally translate into reflecting particles at the side walls~$\Gamma_N$. Furthermore, we have $T=0$ at the top and $T=1$ for the temperature in dimensionless units. These conditions simply translate into absorbing every trajectory hitting the top lid, and re-injecting a new one at a random position at the bottom lid. Trajectories are reflected once they try to exit at the bottom lid. Note that the dynamical equation~\eqref{eq:SDE_Langevin} is independent of the choice of a reference temperature, which is well aligned with the intuition that the dynamics of single heat packages should not depend on the reference temperature. The total heat transport depends on the reference temperature through the initial distribution~$T_0$. The absorption and re-injection of trajectories at the boundaries results in some trajectories with a shorter life span than the considered total integration time since they leave before the end or are seeded later.

Combining \cite[Theorem~3]{Banisch2017} with~\cite[Sec. 4.1]{Froyland2014} shows that eigenvectors of the newly obtained matrix~$\Pdm$ are relaxed solutions of the problem  (cf.\ \eqref{eq:coh_quant})
\begin{equation} \label{eq:coherenceproblem}
\max_{\mathbb{A}} \frac{T(\mathbb{A}\to \mathbb{A})}{T(\mathbb{A})} +  \frac{T(\mathbb{A}^C\to \mathbb{A}^C)}{T(\mathbb{A}^C)},
\end{equation}
where $\mathbb{A} \subset \{1,\ldots, N\}$ is a subset of all trajectories (including trajectories seeded later), $\mathbb{A}^C$ is its complement, $T(\mathbb{A})$ is the total heat content of the set $\mathbb{A}$ (as every particle is a heat package with one unit of heat, this is the cardinality of~$\mathbb{A}$), and $T(\mathbb{A}\to \mathbb{B})$ is the heat moving from set $\mathbb{A}$ to set $\mathbb{B}$ under the heat dynamics. Thus, $T(\mathbb{A}\to \mathbb{A})$ describes the heat remaining in $\mathbb{A}$. Solutions of~\eqref{eq:coherenceproblem} attempt to partition the domain into two subdomains between which there is as little heat exchange as possible, while the normalization by~$T(\mathbb{A})$ and~$T(\mathbb{A}^C)$ avoids highly unbalanced partitions where one of the sets contains almost all of the heat.

While the problem \eqref{eq:coherenceproblem} is clearly not invariant under translation with respect to a reference temperature, in certain cases it can be argued that it would give highly consistent solutions for any reference value. Let us assume that both $T(\mathbb{A})\approx T(\mathbb{A}^C)$, and that the trajectories in $\mathbb{A}$ and $\mathbb{A}^C$ cover approximately the same area in physical space. Then problem~\eqref{eq:coherenceproblem} and the problem~$\min_{\mathbb{A}} \{T(\mathbb{A}\to \mathbb{A}^C) + T(\mathbb{A}^C\to \mathbb{A})\}$ have approximately the same solutions, while the latter describes total heat exchange between the two subdomains. Now, the physically relevant quantity is the \emph{net heat flow}, which is the difference of $T(\mathbb{A}\to \mathbb{A}^C)$ and $T(\mathbb{A}^C\to \mathbb{A})$, and this is almost invariant under changing the reference measure, if the spatial domains occupied by $\mathbb{A}$ and $\mathbb{A}^C$ are almost of the same size, because the contributions due to translation by a reference value cancels out.

\subsection{Heat coherence in two-dimensional Rayleigh--B\'enard convection}
\begin{figure}
\begin{center}
\includegraphics[width=0.4\textwidth]{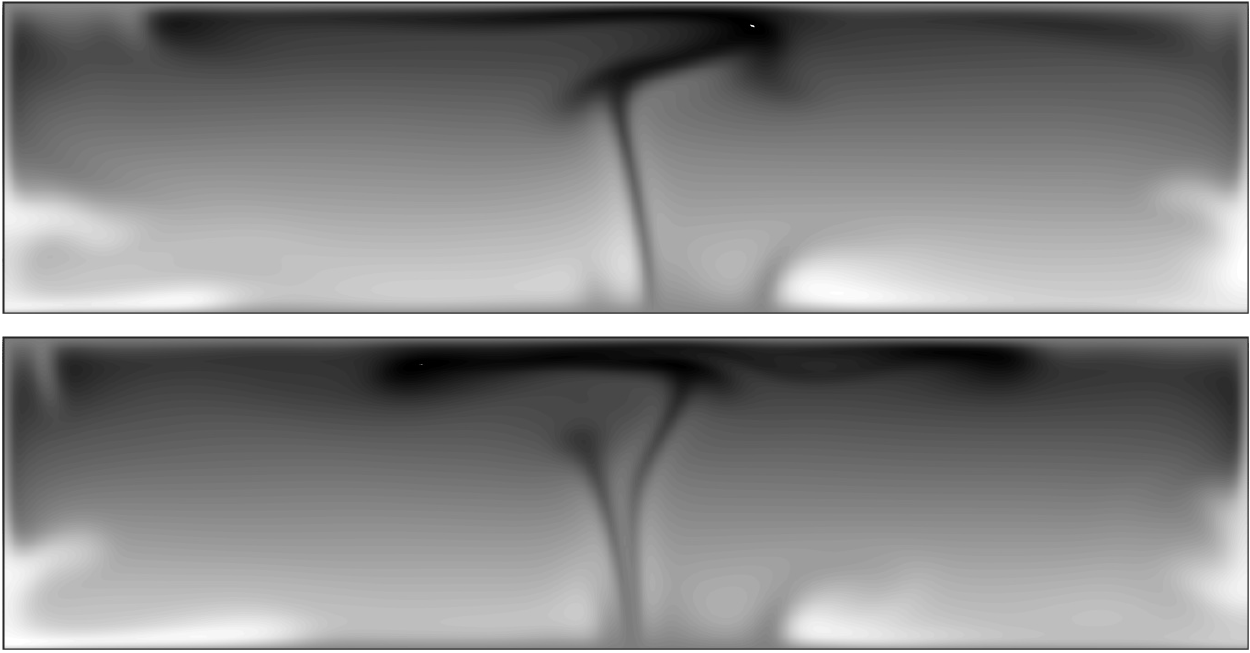}\\
\caption{Temperature contour gray scale plot in relation to the heat coherence evaluation. Initial slice (top) at $t=2000 t_f$ and final slice (bottom) at $t=2010 t_f$. Dimensionless temperature varies between 0 (bright) and 1(dark).}
\label{fig:TempSlice}
\end{center}
\end{figure}

\begin{figure}
\begin{center}
\includegraphics[width=0.4\textwidth]{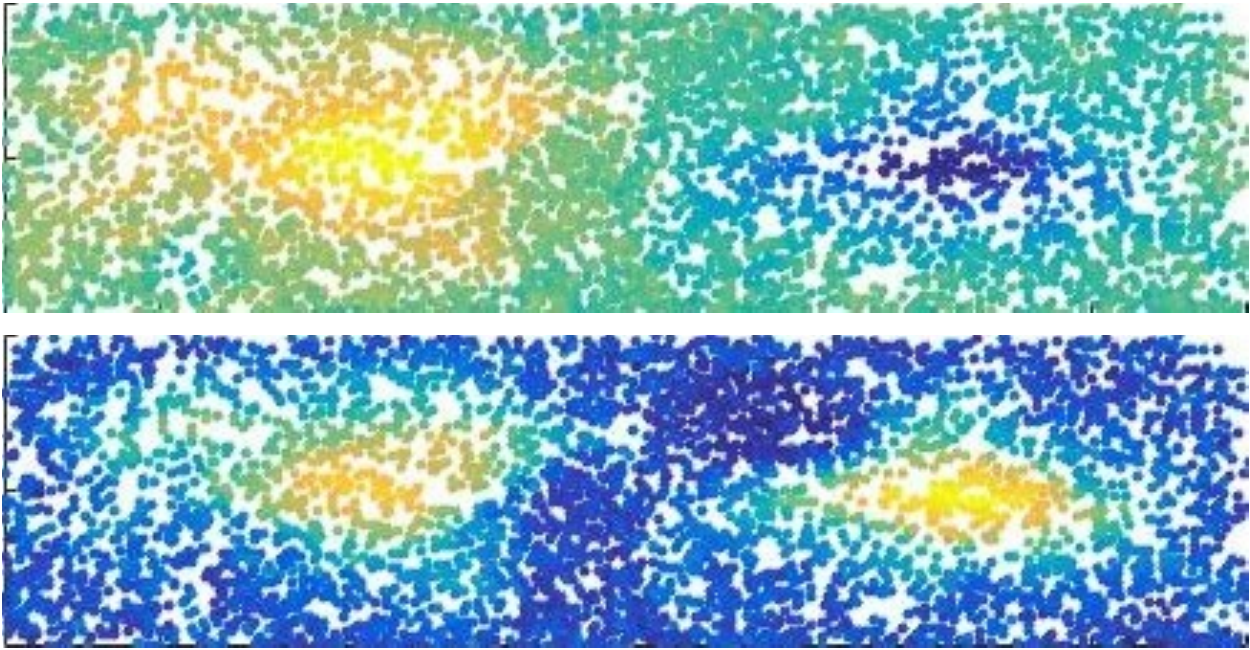}\\
\caption{Long-term heat coherence analysis in two-dimensional Rayleigh--B\'{e}nard flow. Second eigenvector (top) and third eigenvector (top) are shown. Negative (positive) values are indicated by dark (bright) colors.}
\label{fig:HeatLongEV}
\end{center}
\end{figure}

In the following, this methodology is applied to the present two-dimensional Rayleigh--B\'enard system. We conduct the analysis for the diffusion maps method (\ref{sec:dm}) only since the other two methods gave similar results as we saw in Sec.~\ref{sec:comparison}. We first look again at a small time interval which means here $\mathcal{T} = 100$ steps which corresponds to a time interval of $10 t_f$.
\begin{figure}
\centering
\includegraphics[width=0.4\textwidth]{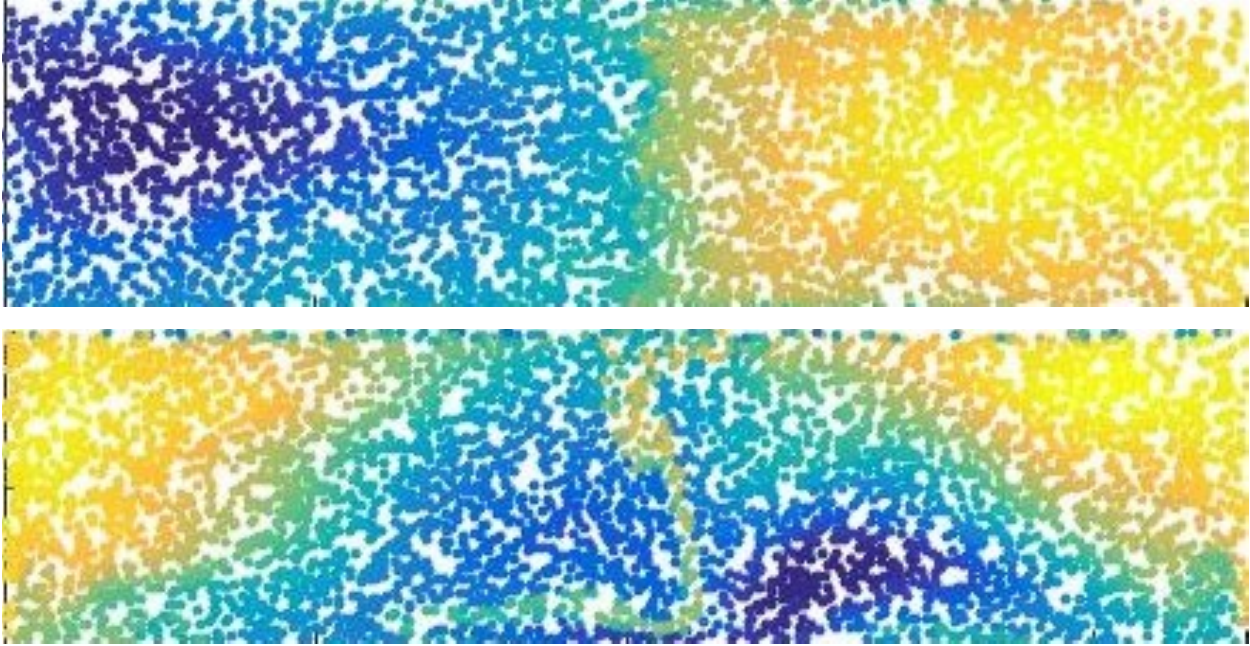}
\caption{Short-term heat coherence analysis in two-dimensional Rayleigh--B\'{e}nard flow. Second eigenvector (top) and third eigenvector (bottom) are shown. Negative (positive) values are indicated by dark (bright) colors. Eigenvectors of $\Qdm$ at the initial time.}
\label{fig:HeatEV-1}
\end{figure}
If we look at the initial and the final time slices (see Figs. \ref{fig:HeatEV-1} and \ref{fig:HeatEV-3}, respectively) we can see a separation of the right and left sides of the domain. Furthermore, the color of the particles that leave the domain fast and particles that enter the domain late is green, which implies that they have a zero value in the eigenvector. These Lagrangian particles correspond to thermal plumes as seen in the temperature field snapshot in Figure \ref{fig:TempSlice}.
\begin{figure}
\begin{center}
\includegraphics[width=0.4\textwidth]{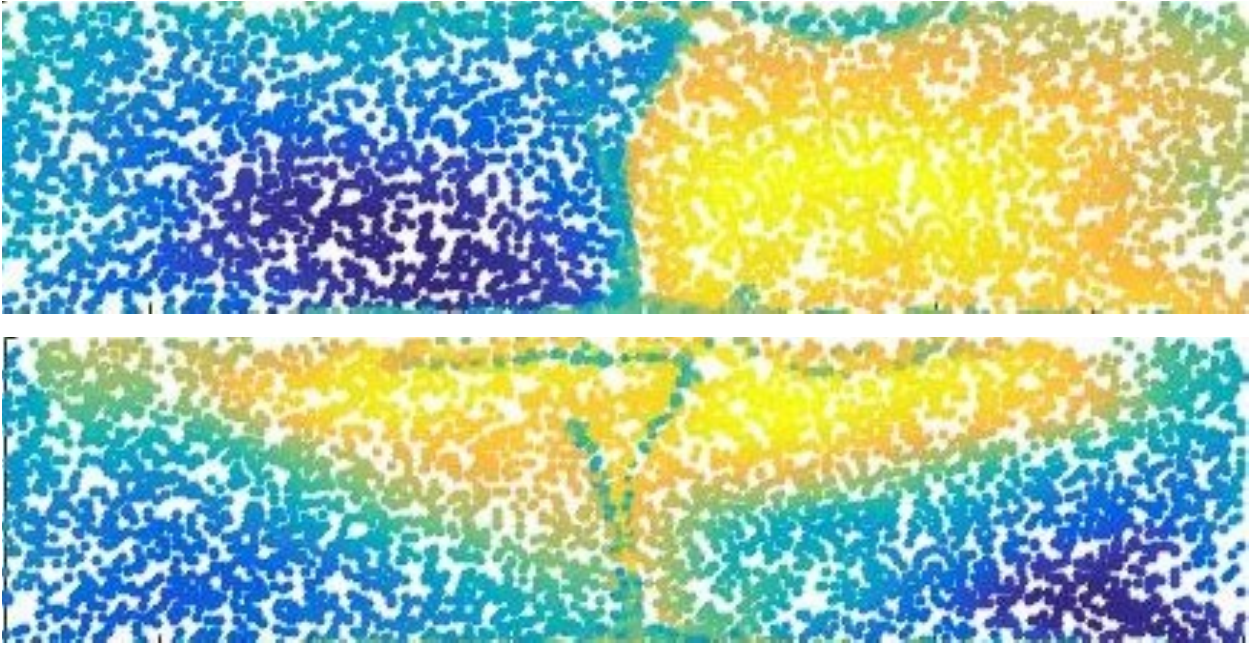}
\caption{Same as Figure \ref{fig:HeatEV-1}, but for the final time of the short time interval.}
\label{fig:HeatEV-3}
\end{center}
\end{figure}
The algorithm identifies them as less relevant for the overall dynamics due to their short life span, which in turn implies that they are highly relevant for the effective heat transport from the bottom to the top.
We now look at a longer sequence of $\mathcal{T} = 2000$ steps which corresponds to a time span of $200 t_f$. The second eigenvector (see Figure \ref{fig:HeatLongEV}, top panel) reveals now that the cores of the convection rolls are the most coherent features for the temperature evolution. This implies that there is almost no effective heat transport through the cores aside from diffusion which will become increasingly subdominant as the Rayleigh number grows.
It is important to note that even though the most heat-coherent sets do not contribute to vertical heat transport in this example, other cases may arise as well. If there would be big bubbles of hot fluid moving (slowly, compared to the considered time span) from the bottom to the top, the algorithm would identify them as heat coherent. 

\section{Conclusion}
\label{sec:conclusion}

The main objectives of this work were to discuss coherence in a simple two-dimensional turbulent convection flow from a Lagrangian point of view and to relate it to the more frequently used perspective of the Eulerian frame of reference. We therefore compared three different Lagrangian approaches, (i) minimum distance spectral-clustering based, (ii)  diffusion-map based, and (iii) dynamic Laplacian based analysis of Lagrangian data.  We find that all three methods identify similar coherent sets, the core regions of the two convection rolls in our example flow. These are the areas in which neighboring Lagrangian particles would remain close to each other for the longest time before dispersion by small-scale turbulence would tear them apart from each other. Our analysis shows that these regions are the complementary to those which would be highlighted in a time-averaging procedure in the Eulerian frame of reference, namely the ridges of hot upwelling and cold downwelling fluid between these two circulation rolls. However, the notion of coherence is less strict in the minimum distance spectral-clustering based analysis compared to the diffusion-map based analysis and it varies with interval length for the dynamic Laplacian based analysis.

Furthermore, we introduced the concept of time averaging in the Lagrangian frame and demonstrated that -- similar to the Eulerian case, coherence of structures is improved. 

Finally, we discussed the concept of heat coherence in the present setting. We therefore suggested an approach to analyze the transport of non-passive scalar quantities including boundary sources and sinks utilizing the theory of coherent sets. We find that effective heat transport only occurs outside the cores.

How do the present Lagrangian and the (standard) Eulerian description compare to each other? In RBC flows, the prominent structures are flow circulations between the top and bottom plates in connection with rising hot or falling cold thermal plumes. The Eulerian picture in Figure \ref{fig:field} reveals a space-filling coherent pattern with a characteristic length $\lambda$ (which is here simply the horizontal extension of a pair of counter-rotating rolls) and a characteristic time scale $\tau$ which is proportional to the typical turnover time inside a roll \cite{Pandey:NatCommun2018}. The Eulerian analysis highlights the regions of the flow that contribute most to the convective heat transfer, namely, the hot upwellings and the cold downwellings between the counter-rotating rolls. Lagrangian methods are connected to the material transport by the evolving flow. As shown in this work, all Lagrangian methods detect the core regions of the large-scale circulation rolls as the coherent sets in which fluid particles remain together for the longest time. Thus, they reveal regions that form the spatial complement to the Eulerian ones, those that contribute least to the turbulent heat transfer. One major advantage of the discussed Lagrangian analysis methods is their objectivity, i.e., coordinate-frame independence (see, e.g.,\ \cite{Haller2015}) that helps to identify barriers to the turbulent transport in the flow. Lagrangian methods provide complementary powerful tools to reduce the dynamics to that of a few relevant degrees of freedom. They may not only be applicable in numerical simulations, but also for the growing number of experimental techniques that provide Lagrangian particle tracks, e.g. \cite{Monaro2019,Cierpka2013,Schanz2016}.

The present example was a two-dimensional flow at a moderate Rayleigh number in a working fluid with a large Prandtl number. Such a setup is an appropriate starting point for Lagrangian studies as the temperature field obeys a small diffusivity and the magnitude of turbulent velocity fluctuations remains small. The Reynolds number which quantifies the turbulent momentum transfer is ${\rm Re}\approx 96$ in our case. As a part of the future work, we will extend the mathematical foundations of the present Lagrangian framework to temporally averaged turbulence fields and apply these  techniques in three-dimensional settings for extended flows. This is necessary to compensate for the enhanced turbulent dispersion which is always probed by Lagrangian methods and which increases as Rayleigh numbers are increased or Prandtl numbers are decreased. These efforts are partly under way and will be reported elsewhere.

\section*{Acknowledgments}
This work of C.S., M.S. and A.P. is supported by the Priority Programme DFG-SPP 1881 ''Turbulent Superstructures'' of the Deutsche Forschungsgemeinschaft (DFG). We acknowledge computing resources by the Large Scale Project pr62se of the Gauss Centre for Supercomputing. The authors thank Christian Cierpka, Daniel Karrasch and Nathanael Schilling for helpful comments.

\appendix
\section{Additional Material}
\subsection{Interpretation in terms of graph Laplacians.}
\label{app:dynLap_graphLap}

The stiffness matrices $K^k$ are symmetric and have zero row and column sums \cite{FroylandJunge2018}. Its diagonal entries are positive and the off-diagonal entries non-positive in certain important cases (and also in all numerical experiments).  Because of this, for the graph $G^k$ with nodes $X^k$ and edges defined by the edges of the triangulation, we can write $K^k = \Pi^k-W^k$, where
$$W^k_{ij} = \left\{
            \begin{array}{ll}
              -K^k_{ij}, & \hbox{$i\neq j$,} \\
              0, & \hbox{$i=j$,}
            \end{array}
          \right.$$
are the \textit{weights} assigned to edges $(i,j)$, and $\Pi^k_{ii}=K^k_{ii}$ is a diagonal matrix with the $i^{\rm th}$ diagonal element equal to the \textit{degree} of node $i$ in $G^k$, namely $\Pi^k_{ii}=-\sum_j K^k_{ij}$.
Thus, we can view $K^k$ as an (unnormalized) graph Laplacian.

Note that the entry $K^k_{ij}$ will decrease in magnitude as the distance between the associated data points ${\bm x}_i(t_k), {\bm x}_j(t_k)$ increases until the Delaunay edge $(i,j)$ no longer exists in $G^k$. 

When we solve the eigenproblem $\bar{K} v = \lambda M v$, we normalize by the symmetric, nonnegative mass matrix $M$, i.e., based on local area or volume elements that neighboring data points enclose.  This normalization is different to the standard graph Laplacian normalization (which is based on node degree only):  $M$ is not diagonal and the small number of off-diagonal entries of $M$ coincide with the off-diagonal entries of $K^0$ and correspond to arcs $(i,j)$ that are in the graph $G^0$.  In fact, normalizing by the mass matrix \emph{automatically handles nonuniformly distributed data}, because if initial points ${\bm x}^0_i, {\bm x}^0_j$ are far apart the value of $M_{ij}$ will be commensurately larger. 

Note further that since a triangulation is used here, there is no free parameter (like the cutoff radius) to choose and that the method can always yield a decomposition of the entire domain $\Omega$ (more precisely, the convex hull of the data points) into coherent sets.

\subsection{Comparison of methods based on the rate-matrix interpretation}
\label{app:theoretical_comp}

From the point of view of rate matrices in Sec.~\ref{ssec:theoretical_comp}, on the one hand, $\Qnw$ defines a process where every state has holding time~$1$ (because~$\Qnw_{ii}=1$), and the process has equal probabilities~$1/D_{ii}$ to jump to any of its ``neighbors'', where $i$ and~$j$ are neighbors if~$A_{ij}=1$. We observe that the proximity parameter~$\ep$ can also be interpreted as diffusion strength (diffusion coefficient), as the holding times of the Markov process are all one, and the ``jumps'' cover an $\ep$-neighborhood in space. This can be seen from noting that a Brownian diffusion~$c\,B_t$ of strength~$c$ has standard deviation~$\sqrt{\mathrm{Var}[c\,B_t]} = c\sqrt{t}$, i.e., if it has strength~$c=\ep$, then it produces~$\ep$ mean deviation in unit time. Thus, in a first order approximation, we could interpret~$\Qnw$ as a diffusion on trajectories with strength~$\ep$.

	On the other hand, $\Qdm$ defines a random walk where the $i$-th state has holding time
	\begin{equation} \label{eq:dm_holdingtime}
		\frac{\delta}{1-\frac{1}{\mathcal{T}}\sum_k \frac{1}{\sum_j K_{t_k,ij}}},
	\end{equation}
	which follows from the construction noting that~$K_{t_k,ii} = 1$. We observe immediately that the scale parameter~$\delta$ features directly as a timescale. As the ``range'' of the kernel \footnote{The cutoff radius~$r$ is defined by~$\smash{k_{\delta}({\bm x},{\bm y}) = h(\rho^2/\delta) < \vartheta}$ for all $\|{\bm x}-{\bm y}\| = \rho > r$, with some small threshold parameter~$\vartheta$. This directly yields~$r = \mathrm{const}\cdot\sqrt{\delta}$.} and thus the mean jump distance is~$\mathcal{O}(\sqrt{\delta})$, a similar consideration as above shows that~$\Qdm$ can be interpreted as a diffusion of strength~$1$.
	Returning to~\eqref{eq:dm_holdingtime}, we see that the holding time grows very large if the trajectory~$i$ has only few and distant neighbors (i.e., it is unlikely to jump over to trajectories that are not alike), and approaches~$\delta$ from above if the trajectory has many close neighbors. The jump probabilities are readily encoded in the entries of~$\Pdm$, and due to the construction involving the similarity matrix, it is more likely that the process jumps to a neighbor which is closer on average (in time).

	To summarize this theoretical comparison of the methods from Secs. \ref{sec:nw} and \ref{sec:dm}, both use spectral clustering with matrices that are interpretable as rate matrices of certain Markov jump processes on the trajectories. By this, those trajectories belong to the same coherent set which are likely to be reached from one another; opposed to unlikely transitions from other trajectories. This behavior is often referred to as \emph{metastability} or \emph{almost invariance}, and its connection to the dominant eigenmodes of the process' jump matrix is well analyzed~\cite{Davies1982a,Davies1982b,Bovier2002,Froyland2005,Huisinga2006,Schuette2013}. The difference in the methods considered here relies in characteristics of the associated Markov processes: The network-based process jumps to its neighbors with equal probabilities (thus utilizing only a binary information about distances; whether they are smaller or larger than $\ep$); while the diffusion-maps based process prefers to jump to closer neighbors (thus utilizing a more refined distance information). While the proximity parameter~$\ep$ can be viewed as strength of the Markov process (diffusion) generated on the trajectories by~$\Qnw$, the strength of the analogous diffusion generated by~$\Qdm$ is always one.
	As these diffusion processes are discrete in space (because of jumping between a finite set of trajectories), their mean jump distance governs the finest scales they can resolve: $\mathcal{O}(\ep)$ and~$\mathcal{O}(\sqrt{\delta})$.  Coherent sets below these scales can not be detected.


\begin{thebibliography}{16}
\expandafter\ifx\csname natexlab\endcsname\relax\def\natexlab#1{#1}\fi
\expandafter\ifx\csname bibnamefont\endcsname\relax
  \def\bibnamefont#1{#1}\fi
\expandafter\ifx\csname bibfnamefont\endcsname\relax
  \def\bibfnamefont#1{#1}\fi
\expandafter\ifx\csname citenamefont\endcsname\relax
  \def\citenamefont#1{#1}\fi
\expandafter\ifx\csname url\endcsname\relax
  \def\url#1{\texttt{#1}}\fi
\expandafter\ifx\csname urlprefix\endcsname\relax\def\urlprefix{URL }\fi
\providecommand{\bibinfo}[2]{#2}
\providecommand{\eprint}[2][]{\url{#2}}

\bibitem{Miesch2005}
M. S. Miesch, Living Rev. Solar Phys. {\bf 2}, 1 (2005).

\bibitem{Christensen2006}
U. R. Christensen and J. Aubert, Geophys. J. Int. {\bf 166}(1), 79 (2006).

\bibitem{Stevens2005}
B. Stevens, Annu. Rev. Earth Planet. Sci. {\bf 33}, 605 (2005).

\bibitem{Bouffard2019}
D. Bouffard and A. Wuest, Annu. Rev. Fluid Mech. {\bf 51}, 189 (2019). 

\bibitem{Chilla:EPJE2012}
F.~Chill\`{a} and J. Schumacher, Eur. Phys. J. E {\bf 35}, 58 (2012).

\bibitem{Hartlep2003}
T. Hartlep, A. Tilgner, and F. H. Busse, Phys. Rev. Lett. {\bf 91}, 064501 (2003).

\bibitem{Hardenberg2008}
J. von Hardenberg, A. Parodi, G. Passoni, A. Provenzale, and E. A. Spiegel, Phys. Lett. A {\bf 372}, 2223 (2008).
 
\bibitem{Bailon2010} 
J. Bailon-Cuba, M. S. Emran, and J. Schumacher,  J. Fluid Mech. {\bf 655}, 152 (2010). 

\bibitem{Emran2015}
M. S. Emran and J. Schumacher, J. Fluid Mech. {\bf 776}, 96 (2015).  

\bibitem{Stevens2018}
R. J. A. M. Stevens, A. Blass, X. Zhu, R. Verzicco, and D. Lohse, Phys. Rev. Fluids {\bf 3}, 041501(R) (2018).

\bibitem{Pandey:NatCommun2018}
A. Pandey, J. D. Scheel, and J. Schumacher, Nat. Commun. {\bf 9}, 2118 (2018).

\bibitem{Bailon2011}
J. Bailon-Cuba and J. Schumacher, Phys. Fluids {\bf 23}, 077101 (2011).

\bibitem{Podvin2012}
B. Podvin and A. Sergent, Phys. Fluids {\bf 24}, 105106 (2012).

\bibitem{Verma2017}
S. Paul and M. K. Verma, {\em Proper Orthogonal Decomposition vs. Fourier analysis for extraction of large-scale structures of thermal convection}, Proceedings of "Advances in Computation, Modeling and Control of Transitional and Turbulent Flows", pp. 433--441 (2017).
 
\bibitem{Verma2018}
M. K. Verma, {\em Physics of Buoyant Flows}, World Scientific, Singapore, 2018.

\bibitem{FrLlSa10}
G. Froyland, S. Lloyd, and N. Santitissadeekorn. Physica D, 239: 1527-1541 (2010).

\bibitem{Froyland2013}
G. Froyland, Physica D {\bf 250}, 1 (2013).

\bibitem{Allshouse2015}
M. R. Allshouse and T. Peacock, Chaos {\bf 25}, 097617 (2015).

\bibitem{Karrasch2017}
D. Karrasch and J. Keller, arXiv:1608.05598 (2016).

\bibitem{DellnitzFroylandJunge2001}
M. Dellnitz, G. Froyland and O. Junge, in: Ergodic theory, analysis, and efficient simulation of dynamical systems, 145-174, Springer (2001). 

\bibitem{DellnitzJunge2002}
M. Dellnitz and O. Junge, Set oriented numerical methods for dynamical systems. Handbook of dynamical systems Vol.2, 221-264, North-Holland, Amsterdam (2002).

\bibitem{Hadjighasem2017}
A. Hadjighasem, M. Farazmand, D. Blazevski, G. Froyland, and G. Haller,
Chaos {\bf 27}, 053104 (2017).

\bibitem{Haller2015}
G. Haller, Annu. Rev. Fluid Mech. {\bf 47}, 137 (2015).

\bibitem{Haller2018}
G. Haller, D. Karrasch, and F. Kogelbauer, Proc. Natl. Acad. Sci. USA {\bf 115},  9074 (2018).

\bibitem{Froyland2015}
G. Froyland, Nonlinearity {\bf 28}, 3587 (2015).

\bibitem{Banisch2017}
R. Banisch and P. Koltai, Chaos {\bf 27}, 035804 (2017).

\bibitem{Padberg2017} 
K. Padberg-Gehle and C. Schneide, Nonlin. Processes Geophys.  {\bf 24},  661 (2017).

\bibitem{Schneide2018} 
C. Schneide, A. Pandey, K. Padberg-Gehle, and J. Schumacher, Phys. Rev. Fluids {\bf 3},  113501 (2018).

\bibitem{Schlueter2017}
K. L. Schlueter-Kuck and J. O. Dabiri,
J. Fluid Mech. {\bf 811}, 468 (2017).

\bibitem{Froyland_Padberg_2015}
G.~Froyland and K.~Padberg-Gehle,  Chaos {\bf 25}, 087406 (2015).

\bibitem{Hadjighasem2016}
A. Hadjighasem, D. Karrasch, H. Teramoto, and G. Haller, 
Phys. Rev. E {\bf 93}, 063107 (2016).

 \bibitem{FroylandJunge2018}
G. Froyland and O. Junge, SIAM J. Appl. Dyn. Sys. 17, 1891 (2018)

\bibitem{Speetjens2012}
M.F.M. Speetjens, J. Phys.:~Conf. Ser. {\bf 395}, 012033 (2012).

\bibitem{Schmalzl:EPL2004}
J. Schmalzl, M. Breuer, and U. Hansen, Europhys. Lett. {\bf 67}, 390 (2004).

\bibitem{Poel:JFM2013}
E.~P. van der Poel, R.~J. A.~M. Stevens, and D. Lohse, J. Fluid Mech. {\bf 736}, 177 (2013).

\bibitem{Pandey:Pramana2016}
A. Pandey, M. K. Verma, A. G. Chatterjee, and B. Dutta, Pramana - J. Phys. {\bf 87}, 13 (2016).

\bibitem{Petschel:PRE2011}
K. Petschel, M. Wilczek, M. Breuer, R. Friedrich, and U. Hansen, Phys. Rev. E {\bf 84}, 026309 (2011).

\bibitem{Pandey:PRE2018}
A. Pandey, M. K. Verma, and M. Barma, Phys. Rev. E {\bf 98}, 023109 (2018).

\bibitem{Verma:Pramana2013}
M.~K. Verma, A.~G. Chatterjee, K.~S. Reddy, R.~K. Yadav, S. Paul, M. Chandra, and R. Samtaney, Pramana - J. Phys. {\bf 81}, 617 (2013).

\bibitem{Verma:POF2015}
M.~K. Verma, S.~C. Ambhire, and A. Pandey, Phys. Fluids {\bf 27}, 047102 (2015).

\bibitem{Haller2001}
G. Haller, Physica D {\bf 149}, 4 (2001).

\bibitem{Padberg2009}
K. Padberg, B. Thiere, R. Preis, and M. Dellnitz, Commun. Nonlinear Sci. Numer. Simul.  {\bf 14}, 12 (2009).

\bibitem{Donner2009}
R.V. Donner, S. Barbosa, J.F. Kurths, and N. Marwan, Eur. Phys. J. Spec. Top. {\bf 174}, 1 (2009).

\bibitem{Banisch2019}
R. Banisch, P. Koltai, and K. Padberg-Gehle, to appear in Chaos, arXiv:1901.00799 (2019).

\bibitem{ShiMalik2000}
J. Shi and J. Malik, IEEE Trans. Pattern Anal. Mach. Intell. {\bf 22}, 888 (2000).

\bibitem{Lloyd1982}
S. P. Lloyd, IEEE Inf. Theory {\bf 28}, 2 (1982).

\bibitem{Coifman2006}
R.R. Coifman and S. Lafon, Appl. Comput. Harmonic Anal. {\bf 21}, 5 (2006).

\bibitem{FroylandKwok2017}
G. Froyland and E. Kwok, J. Nonlinear Sci., pp.1-83 (2017).

\bibitem{Thiffeault2004}
J.-L Thiffeault, C. R. Doering, and J. D. Gibbon, J. Fluid Mech. {\bf 521}, 105 (2004). 

\bibitem{Dellnitz1999}
M. Dellnitz and O Junge, SIAM J. Numer. Anal. {\bf 36},  491 (1999).

\bibitem{Denner2017}
A. Denner, Ph.D. thesis, TU Munich, 2017.

\bibitem{Donner2010}
R.V. Donner, Y. Zou, J.F. Donges, N. Marwan, and J. Kurths, Phys. Rev. E {\bf 81}, 015101(R) (2010).

\bibitem{Petschel2016}
K. Petschel, S. Stellmach, M. Wilczek, J. L\"ulff, and U. Hansen, Phys. Rev. Lett. {\bf 110}, 114502 (2013).
 
\bibitem{Kimura1983}
S. Kimura and A. Bejan, J. Heat Transfer {\bf 105}, 916 (1983).

\bibitem{Costa2006} 
V.A. Costa, Appl. Mech. Rev. {\bf 59}, 126 (2006).

\bibitem{Pratt2016}
L. Pratt, R. Barkan, and I. Rypina, Fluids {\bf 1}, 27 (2016).

\bibitem{Balasuriya2018}
S. Balasuriya, N.T. Ouellette, and I. Rypina, Physica D {\bf 372}, 31 (2018).

\bibitem{Speetjens2012a}
M.F.M. Speetjens, Int. J. Therm. Sci. {\bf 61}, 79 (2012).

\bibitem{Froyland2014}
G. Froyland and K. Padberg-Gehle, in {\em Ergodic Theory, Open Dynamics, and Coherent Structures. PROMS, vol. 70}, edited by W. Bahsoun {\em et al.} (Springer, New York, 2014).

\bibitem{Monaro2019}
M. Novara, D. Schanz, R. Geisler, S. Gesemann, C. Voss, and A. Schr\"oder, Exp. Fluids {\bf 60}, 44 (2019).

\bibitem{Cierpka2013}
C. Cierpka, B. L\"utke, and C. J. K\"ahler, Exp. Fluids {\bf 54}, 1533 (2013).

\bibitem{Schanz2016}
D. Schanz, S. Gesemann, and A. Schr\"oder, Exp. Fluids {\bf 57}, 70 (2016).

\bibitem{Davies1982a}
E.B. Davies, Proc. Lond. Math. Soc. {\bf 3}, 133 (1982).

\bibitem{Davies1982b} 
E.B. Davies, J. Lond. Math. Soc. {\bf 2}, 541 (1982).

\bibitem{Bovier2002} 
A. Bovier, M. Eckhoff, V. Gayrard, and M. Klein, Comm. Math. Phys. {\bf 228}, 219 (2002).

\bibitem{Froyland2005} 
G. Froyland, Physica D {\bf 200}, 205 (2005).

\bibitem{Huisinga2006} 
W. Huisinga  and B. Schmidt, in {\em  New Algorithms for Macromolecular Simulation},  Lecture Notes in Computational Science and Engineering, vol. 49, edited by B. Leimkuhler {\em et al.} (Springer, Berlin, Heidelberg, 2016).

\bibitem{Schuette2013}
C. Sch\"utte and M. Sarich, {\em Metastability and Markov State Models in Molecular Dynamics}, Courant Lecture Notes in Mathematics, vol. 24 (AMS, Providence, 2013).


\end{thebibliography}
\end{document}